\documentclass[10pt]{article}
\usepackage{mathrsfs}
\usepackage{amsmath,amsfonts,amssymb,mathtools}
\usepackage{cite}
\usepackage{dsfont}
\usepackage{graphicx}
\usepackage{hyperref}
\usepackage{verbatim}
\usepackage{slashed}
\usepackage[all]{xy}
\usepackage{xcolor}
\usepackage{subfig}
\usepackage{tikz}
\usetikzlibrary{shapes.geometric,arrows}
\usepackage{relsize}
\usepackage{caption}
\usepackage{float}
\usetikzlibrary{positioning}
\usepackage{geometry}
\usepackage[T1]{fontenc}
\usepackage{lmodern}
\usepackage{stackengine} %per \ocirc 
 
\usepackage[utf8]{inputenc}

\hypersetup{
  colorlinks,
  citecolor=violet,
  linkcolor=blue,
  urlcolor=blue}

\csname @addtoreset\endcsname{equation}{section}
\DeclareMathAlphabet\mathbfcal{OMS}{cmsy}{b}{n}

\newcommand{\beq}{\begin{equation}}
\newcommand{\eeq}{\end{equation}}
\newcommand{\bea}{\begin{eqnarray}}
\newcommand{\eea}{\end{eqnarray}}
\newcommand{\ba}{\begin{array}}
\newcommand{\ea}{\end{array}}
\newcommand{\bit}{\begin{itemize}}
\newcommand{\eit}{\end{itemize}}
\newcommand{\nn}{\nonumber}

\newcommand{\mezzo}{\frac{1}{2}}
\newcommand{\complesso}{{\ \hbox{{\rm I}\kern-.6em\hbox{\bf C}}}}
\newcommand{\reale}{{\hbox{{\rm I}\kern-.2em\hbox{\rm R}}}}
\newcommand{\uno}{ \,  \raisebox{+0.14em}{{\hbox{{\rm \scriptsize ]}} \raisebox{-0.2em}{\kern-.8em\hbox{1}}}} \, }  %  operatore identit\`a

\newcommand{\p}{\partial}
\renewcommand{\a}{\alpha}
\renewcommand{\b}{\beta}
\newcommand{\g}{\gamma}

\newcommand{\D}{\Delta}
\newcommand{\e}{\epsilon}
\newcommand{\Er}{{\mathbfcal{E}}}

\renewcommand{\k}{\kappa}
\renewcommand{\l}{\lambda}
\renewcommand{\L}{\Lambda}

\newcommand{\m}{\mu}

\newcommand{\n}{\nu}
\renewcommand{\r}{\rho}
\newcommand{\s}{\sigma}

\renewcommand{\S}{\Sigma}

\newcommand{\om}{\omega}
\newcommand{\Om}{\Omega}

\begin{document}

%\begin{comment}

\begin{titlepage}

\vspace{0.3cm}

\begin{flushright}
%$IFUM$--1105--$FT$ \\
$LIFT$--13-1.26
\end{flushright}

\vspace{0.7cm}

\begin{center}
\renewcommand{\thefootnote}{\fnsymbol{footnote}}
\vskip 1mm  
{\Huge \bf             Black holes in rotating, electro-  \\
\vskip 5mm
                      magnetic backgrounds and topological \\
\vskip 8mm
                           Kerr-Newman-NUT spacetimes
}
\vskip 28mm
%Black holes in rotating, electromagnetic backgrounds and topological Kerr-Newman-NUT spacetime
%Black holes in the external Bertotti-Robinson-Bonnor-Melvin electromagnetic field
%Rotating and swirling binary black hole system balanced by its gravitational spin-spin interaction
%Most general Type-D Black Hole and the Accelerating Reissner-Nordstrom-NUT-(A)dS solution
%Ultrarelativistic boost of a black hole in the magnetic universe of Levi-Civita--Bertotti--Robinson

{\large {Marco Astorino$^{a}$\footnote{marco.astorino@gmail.com}
% Matilde Torresan$^{b}$\footnote{matilde.torresan@studenti.unimi.it}
}}\\
\renewcommand{\thefootnote}{\arabic{footnote}}
\setcounter{footnote}{0}
\vskip 8mm
\vspace{0.2 cm}
{\small \textit{$^{a}$Laboratorio Italiano di Fisica Teorica (LIFT),  \\
Via Archimede 20, I-20129 Milano, Italy}\\
} \vspace{0.2 cm}
%{\small \textit{$^{b}$Istituto Nazionale di Fisica Nucleare (INFN), Sezione di Milano \\
%Via Celoria 16, I-20133 Milano, Italy}\\
%} 
%\vspace{0.2 cm}
%
%{\small \textit{$^{b}$Universit\`a degli Studi di Milano}} \\
%{\small {\it Via Celoria 16, I-20133 Milano, Italy}\\}

\end{center}

\vspace{1.8cm}

\begin{center}
{\bf Abstract}
\end{center}
{We observe that a large class of well behaved stationary and axisymmetric black hole solutions in general relativity and in the Einstein-Maxwell theory can be classified according to the properties of their background. Indeed all these backgrounds belong to a unique family which includes simultaneously all the known axisymmetric and regular backgrounds: the swirling, the Bertotti-Robinson, the Bonnor-Melvin universe, Witten's expanding bubble and also other novel, regular, rotating gravitational or electromagnetic environments. All these can be, fundamentally, traced back to the double Wick rotation of the topological generalisation of (accelerating) Kerr-Newman-NUT metric.\\
We present a black hole embedded in an unexplored sector of the general background: Schwarzschild inside a generalised rotating (and possibly electromagnetic) universe. \\
These results indicate that basically all the known analytical and exact single black hole solutions in the four-dimensional Einstein-Maxwell theory belong to the (accelerating) Kerr-Newman-NUT family embedded into backgrounds that are a subcase of the conjugated Kerr-Newman-NUT space-time with an angular manifold of arbitrary topology.}

\end{titlepage}

\addtocounter{page}{1}

\newpage

%\tableofcontents
%\newpage

\section{Introduction}
\label{sec:introduction}

Black holes in vacuum general relativity are very peculiar objects because they are the most restricted and constrained solutions. Actually when  asymptotic flatness is assumed the only well defined\footnote{We are referring only to analytical and exact solution of the vacuum or electro-vacuum Einstein equations, which can be interpreted as black holes, therefore without singularities outside the event horizon.} black hole is the Kerr one \cite{kerr} (which, in the static case, reduces to the Schwarzschild black hole). In the presence of the Maxwell fields it is possible to find its electromagnetic extension, the Kerr-Newman black hole \cite{newman}, but nothing else. \\
On the other hand, relaxing the asymptotic flatness, a few more black hole solutions are known in pure general relativity, for instance the Kerr black hole embedded into a swirling background \cite{swirling}, \cite{removal} or the Kerr black hole embedded into a bubble of nothing \cite{kerr+bubble}. A few other possibilities can be found if we include the cosmological constant in the action or if gravity is coupled to other forces; however in this article we are not taking into account any energy momentum tensor apart from the Maxwell one. \\
For instance, considering the electromagnetic contribution to the Einstein equations, while neglecting the asymptotic flatness condition, the Kerr-Newman-Melvin \cite{ernst-wild} or the Kerr-Bertotti-Robinson spacetimes \cite{kerr-bertotti} or a combination of both \cite{bh+BRBM} can be included in this list. These spacetimes describe black holes embedded in some external back-reacting electromagnetic backgrounds.\\ 
However note that all these asymptotically non-constant curvature generalisations of the Kerr black hole can hardly be considered completely different black holes with respect to the Kerr family. They can be better physically interpreted as continuous deformation of the Kerr(-Newman-NUT) black hole class because they are essentially the Kerr-Newman-NUT black hole embedded in some external background. In fact the Kerr black hole can always be retrieved in some limit, basically vanishing the background. This is in line with the spirit of the uniqueness theorems for axisymmetric and stationary metrics \cite{uniqueness}. \\ 
In this paper we would like to extend possible backgrounds for embedding black holes by introducing a new rotational feature in the asymptotic geometry with respect to the one of the {\it swirling universe}. From a mathematical point of view we know (see \cite{swirling}, for details \cite{malek}) that the swirling background can be obtained from a double Wick rotation of a Taub-NUT metric with a flat base manifold. At the same time it has been shown in \cite{charging} that the Bonnor-Melvin universe (with or without the cosmological constant \cite{charging}, \cite{bonnor}-\cite{melvin}) can be obtained again from a double Wick rotation of the Reissner-Nordstrom solution (with a flat angular surface). Thus it is natural to ask which is the double Wick rotation of the Kerr black hole, or more in general of the Kerr-Newman-NUT black hole. If the swirling background and the external electromagnetic Bonnor-Melvin-Bertotti-Robinson field are respectively related to the conjugated NUT parameter and monopolar electromagnetic charges, what feature is introduced in the metric by the conjugated angular momentum?  We will focus specifically on the possibility that these backgrounds can be used as adequate environment to embed black hole solutions\footnote{This point was preliminarily addressed in \cite{tesi-omar}, while a first study about possible novel backgrounds has been carried out in \cite{tesi-riccardo}.}. We will remain as general as possible to construct possible backgrounds, thus not limiting to flat base manifold black holes' metrics, because, as shown in \cite{bh+BRBM} or in \cite{bubble}, a curved geometry of the base manifold can be necessary for including the Bertotti-Robinson or the expanding bubble of nothings backgrounds. Whereas all these backgrounds admit the presence of the cosmological constant, the lack of theoretical tools to deal with solution generating techniques with $\L$, does not allow us to exploit these backgrounds to accommodate black holes. In fact, as pointed out in \cite{charging}, it is not apparent if in the presence of the cosmological constant axisymmetric and stationary geometries enjoy integrability properties in general relativity, outside the Petrov-D type subclass. \\
Other possible generalisations of these type D backgrounds might be obtained through solution generating techniques. However, at the moment, it is not clear if they can be suitable to host legitimate black holes or if they can carry some more fundamental physical properties, other than being just a composition of known elementary backgrounds.\\

\section{Generalised electromagnetic and rotating backgrounds}
\label{sec:backgrounds}

\subsection{The topological Kerr-Newman-NUT solution}
Consider the Kerr-Newman metric with a cosmological constant and an arbitrary curvature in the angular\footnote{We are mainly using, here, spherical-like coordinates ($t, r, x, \varphi$), where the polar angle is related to $x$ by $x = \cos \vartheta $.} section of the metric, i.e. for constant time and radial coordinates ($t,r$)
\beq \label{KN-topo}
          ds^2 = -\frac{\D_r}{\S^2} \left[d\tau-(2\ell x + ax^2-\omega_0) d\varphi \right]^2 + \frac{\S^2}{\D_r} dr^2 + \frac{\S^2}{\D_x} dx^2 + \frac{\D_x}{\S^2} \left[(r^2+\ell^2+a\omega_0)d\varphi+a d\tau \right]^2 \,,
\eeq
where
\bea
         \D_r(r)  &=& e^2 + p^2 - k\ell^2 + \L\ell^4+a^2[1+k-k^2(1+\ell^2\L)] - 2 m r + (k-a^2k^2\frac{\L}{3}-2\ell^2\L) r^2 - \frac{\L}{3} r^4 \ ,\nn \\
         \D_x(x)  &=& 1+k_7+k(1-k-x^2) + \frac{ak_7x}{\ell} + \frac{ax\L}{3} (k^2-x^2)(ax+4\ell)     \ ,  \nn \\
            \S(r,x) &=& \sqrt{r^2+(\ell+ax)^2}  \ . 
\eea
The associated electromagnetic vector potential is 
\beq\label{KN-Am}
        A_\m = \left[\frac{er-p(\ell+ax)}{\S^2} ,0 \ ,0 \ , \frac{x[p\ell(ax+\ell)-pr^2-er(ax+2\ell)]}{\S^2}+\om_0 A_t(r,x)\right] \ .
\eeq
This metric\footnote{A similar solution was presented in \cite{ortin}, but it seems it was not fulfilling the field equations.}  represents the most general non-accelerating type-D black hole in the theory of general relativity, possibly coupled with standard Maxwell electromagnetism (aligned with the two double-degenerate principal null directions of the Weyl tensor) and eventually also with the cosmological constant $\L$. Thus it is a solution of the standard Einstein-Maxwell-$\L$ field equations
\bea
           & & R_{\m\n} -\mezzo R g_{\m\n} + \L g_{\m\n} = 2 \left( F_{\m\r} F_{\n}^{\ \r} - \frac{1}{4} F_{\r\s} F^{\r\s} \right)  \ ,\\
           & & \p_\m \left(\sqrt{-g} F^{\m\n} \right) = 0 \ .
\eea
The electromagnetic field strength is defined as usual as $F_{\m\n} = \p_\m A_\n - \p_\n A_\m$. The six integrating constants $m,a,\ell,e,p, k$ are related to physical properties of the solution such as the mass, the angular momentum, the NUT parameter, the monopolar electric and magnetic charges and the curvature of the surface defining the event horizon respectively. This latter parameter $k$ defines the geometry and of the event horizon, which can be of spherical topology for $k>0$, flat, cylindrical or toroidal, depending on the identifications for $k=0$, hyperbolic or a torus of genus bigger than 1 for $k<0$. However note that in the presence of rotation the multi-torus construction may not be compatible with the black hole interpretation \cite{klemm-erratum}.  Usually $k$ is considered to take values in \{-1,0,1\}. In that case $k_7$ can be considered zero, to properly reproduce the possible three topologies of the $(x,\varphi)$ section. $k_7$ might become useful in few circumstances, for instance to get easier limit to the Bertotti-Robinson-Bonnor-Melvin background. \\

\subsection{The general background: conjugated topological Kerr-Newman-NUT}
\label{general-background}

We note that all the legitimate black holes solutions ever found in vacuum general relativity, possibly coupled with standard Maxwell electromagnetism, are embedded in a background contained into the conjugate \cite{chandrasekhar}, \cite{swirling}, or equivalently into the double Wick rotation  of the metric (\ref{KN-topo}). We are referring especially to black hole solutions which possess no curvature, nor conical singularities outside the event horizon\footnote{In the literature there are known also black holes deformed by external multipolar expansion, which are reasonable physical model for collapsed stars. In particular they are not singular outside the event horizon, however their curvature scalar invariants grow as the radial coordinate in certain directions, so they can be asymptotically unbounded.  Thus we are not considering here those backgrounds.}. Therefore the study of the general case is significant, since it can unveil a number of other possible good gravitational and electromagnetic  backgrounds to host black holes metrics, which can be enriched by new integrating constants deforming the black hole, thus new possible physical features. These deformations are always continuous modification of the Kerr(-Newman) metric, therefore are always compatible with the actual phenomenology for small values of the new integrating constants. Similarly, for small values of the deforming background parameters the deviation from the constant curvature usual backgrounds, that is Minkowski or (A)dS, can be as small as one needs. \\
So, as a first step, we start with writing the general background as a conjugate solution of the topological Kerr-Newman-NUT solution (\ref{KN-topo})-(\ref{KN-Am}). The conjugate transformation can be achieved most easily by the analytical continuation, known as the double Wick rotation, given by 
\beq \label{dwr}
\tau \to i \phi \ , \qquad \qquad   \varphi \to i t  \ .
\eeq
%by writing a metric into the Lewis-Weyl-Papapetrou form
%\beq
%       ds^2 = -f(d\tau -\omega d\varphi)^2 + \frac{1}{f} \Big[ e^{2\g} (d\rho^2 + dz^2)  + \rho^2d\varphi^2 \Big] \ ,
%\eeq
%and act with the following map on the three functions
In the presence of an electromagnetic field this map has to be combined with an imaginary rotation of the electromagnetic parameters in order to remain with a real vector potential. In the case under consideration, it means that $e\to i \hat{e}, \ p \to i \hat{p}$, in the Maxwell potential and in $\D_r$ (so $\D_x$ and $\S$ remain unchanged). Finally the conjugate solution to the above topological Kerr-Newman-NUT black hole become
\beq \label{KN-dwr}
          ds^2 = \frac{\D_r}{\S^2} \left[d\phi-(2\ell x + ax^2-\omega_0) dt \right]^2 + \frac{\S^2}{\D_r} dr^2 + \frac{\S^2}{\D_x} dx^2 - \frac{\D_x}{\S^2} \left[(r^2+\ell^2+a\omega_0)dt + a d \phi \right]^2 \,,
\eeq
with 
\beq
       \D_r(r)  = - \hat{e}^2 - \hat{p}^2 - k\ell^2 + \L\ell^4+a^2[1+k-k^2(1+\ell^2\L)] - 2 m r + (k-a^2k^2\frac{\L}{3}-2\ell^2\L) r^2 - \frac{\L}{3} r^4 \ ,\nn \\
\eeq
and the potential 1-form
\beq \label{A-dwr}
          A_\m dx^\m = -\left[ \frac{x\left(\hat{p}\ell(ax+\ell)-\hat{p}r^2-\hat{e}r(ax+2\ell)\right)}{\S^2}+\om_0 A_\phi(r,x)  \right]dt \, - \, \left[ \frac{\hat{e}r-\hat{p}(\ell+ax)}{\S^2}  \right] d\phi \ .
\eeq
The transformation (\ref{dwr}) is neither a diffeomorphism nor proper a change of coordinates, therefore, the general background (\ref{KN-dwr})-(\ref{A-dwr})  is physically inequivalent from the Kerr-Newman spacetime. However the double Wick rotation does not change the Petrov class, therefore (\ref{KN-dwr})-(\ref{A-dwr}) inherits the type D from the conjugate spacetime (\ref{KN-topo})-(\ref{KN-Am}). Therefore the Klein-Gordon equation is separable in this background. \\
The easiest example consists in vanishing all the physical charges, for $k=1$, remaining with the Minkowski spacetime, which is mapped by (\ref{dwr}) into itself up to a coordinate transformation. To be more precise it is the flat space-time in accelerating coordinates, that is the Rindler spacetime.\\    

This general background solution (\ref{KN-dwr})-(\ref{A-dwr}) for $k=0$ and $\Lambda=0$, according to the inverse scattering technique \cite{belinski-book}, \cite{alekseev-inverse} is related to a background consisting in a complex soliton, representing a charged and rotating black hole, at infinity of the complex plane of the spectral parameter \cite{belinski-19}. In the Weyl coordinates $(t,\r,z,\varphi)$, the non-null metric components are
\bea
    \label{belinski-g}          g_{tt} &=& \left[ 2 k_2 R -\frac{1}{\k \bar{\k}}  +\k\bar{\k} (\r^2k_1^2-k_2^2)R^2 \right] \D_t^2   \  , \\
                                g_{t\varphi}  &=& \left[ \frac{k_2}{k_2} + \frac{\k\bar{\k}}{k_1}     \right] \D_t \D_\phi \ ,  \nn \\
                                g_{\r\r} &=& g_{zz} \ = \ \frac{c_0 c_1}{\k \bar{\k}} \ , \nn \\
                                g_{\varphi\varphi}  &=& \k \bar{\k} \ \left( \r^2 - \frac{k_2^2}{k_1^2} \right) \D_\phi^2 \ \nn ,
\eea
where $k_i$ and $c_i$ are constants, whereas $\k$ and $R$ are functions defined as follows
\bea
         \k(\r,z) &=& \left[ 4 k_1^2\r^2- 4 k_2 + 1 + i(c_2 + 8 k_1 k_2 z) \right]^{-1}  \ , \nn \\
         R(\r,z)  &=&  16 k_1 c_2 z + 64 k1^2 k_2 z^2 +c_5 \ .
\eea
The electromagnetic potential one-form is 
\beq   \label{belinski-A}            A = \left[ \frac{c_3}{2} - 4k_1z\sin\g + \frac{k_1 R A_t}{\D_\phi} \right] \D_t dt \ + \  \Big[ \textrm{Im} \left(\k^{-1} \right) \sin \g - \text{Re} \left(\k^{-1} \right) \cos \g \Big] \, \frac{ \k \bar{\k} \ \D_\phi}{2k_1} d\phi \, .
\eeq
The change of coordinates to pass from (\ref{KN-dwr})-(\ref{A-dwr}) to (\ref{belinski-g})-(\ref{belinski-A}) is 
\beq
         r = - \frac{\D_\phi^2 \D_t^2 \r^2 + \hat{e}^2 + \hat{p}^2 -a^2}{m} \ , \qquad x = \frac{\sqrt{c_0c_1}}{\D_t} \,  z 
\eeq
while the parametric map is given by
\bea
          a &=& \frac{8k_1k_2\D_t^2}{\sqrt{c_0 c_1}}  \ , \hspace{1.9cm} m = -  \frac{8k_1^2\D_t^3}{c_0 c_1}   \ , \hspace{1.9cm} \D_\phi = - \frac{8k_1^2 \D_t}{\sqrt{c_0c_1}}    \ , \hspace{1.3cm} \ell = c_2 \D_t \ , \nn \\
      \hat{e}  &=& - \frac{4 k_1\D_t^2}{\sqrt{c_0c_1}} \cos \g \ ,               \hspace{1.3cm}  \hat{p}  = - \frac{4 k_1\D_t^2}{\sqrt{c_0c_1}} \sin \g   \ , \hspace{1.3cm}   \om_0 = - \frac{\sqrt{c_0c_1}}{8 k_1} c_5 \ \ .
\eea 
$c_3$ is proportional to the gauge additive constant of the electric potential, whether present\footnote{In the Kerr-Newman solution above we have not considered for brevity, possibly it can always be added.}. Note that for $k\neq 0$, the metric (\ref{KN-dwr})-(\ref{A-dwr}) is more general and it represents a larger class of backgrounds, particularly suitable to embed black holes in. \\
Let us give below some examples, both for $k=0$ and $k\ne0$, how to recover basically all the regular black hole backgrounds known in the literature at the moment, from the solution (\ref{KN-dwr})-(\ref{A-dwr}). \\

\paragraph{Swirling Bonnor-Melvin, k = 0 }

Considering $k=0=\L$ in (\ref{KN-dwr})-(\ref{A-dwr}) and the following change of coordinates and parameters redefinition\footnote{Note that this choice is not unique.}
\bea
        r &=& -\frac{4\r^2+\hat{e}^2}{2m} \ , \hspace{2.1cm} x \ = \ - \frac{2z}{m} \ , \hspace{3cm}  t \ = \ \frac{\tilde{t}}{2} (b^4+16\jmath^2)^{1/4} \ ,\nn \\
         \phi &=& \frac{-\D_\phi \tilde{\phi}}{(b^4+16\jmath^2)^{1/4}} \ ,        \hspace{1.5cm} m \ = \ \frac{4}{(b^4+16\jmath^2)^{1/4}}  \ , \hspace{1.5cm}  \ell \ = \ \frac{8\jmath}{(b^4+16\jmath^2)^{3/4}} \ ,  \\ 
          \hat{e} &=& \ \frac{4b}{(b^4+16\jmath^2)^{1/2}} , \hspace{1.6cm} \hat{p} \ = \ a \ = \ 0 \ , \hspace{2.45cm}  \om_0 \ = \ 0  \ ,\nn
\eea
we obtain the Bonnor-Melvin universe in a swirling background, originally found by Harrison \cite{harrison}. In a renewed form stemming from symmetry transformations of the Ernst equations, as done in \cite{illy}, the metric reads 
\beq \label{g-swirling-melvin}
      ds^2 = \left[\left(1+\frac{b^2\r^2}{4} \right)^2 + \jmath^2\r^4 \right]\left( - d\tilde{t}^2 + dz^2 + d\r^2 \right) + \frac{\r^2(\D_\phi d\tilde{\phi} -4\jmath z d\tilde{t})^2}{\left(1+\frac{b^2\r^2}{4} \right)^2 + \jmath^2\r^4} ,
\eeq
and its electromagnetic 1-form potential\footnote{Usually, when the Harrison transformation is used to embed a given seed into the swirling Bonnor-Melvin background, an apparently alternative electromagnetic potential is generated: $
A=-\frac{2b(4dtjz-\D_\phi d\phi)\r^2(4+b^2\r^2)}{16+8b^2\r^2+(b^4+16\jmath^2)\r^4}$.
This one-form is not related by gauge transformations to the potential (\ref{A-swirling-melvin}), in fact they produce different electromagnetic Faraday tensors, but the same Maxwell energy momentum tensor. Anyway when $b$ or $\jmath$ vanish the two electromagnetic fields coincide. They are related through a duality rotation of the electromagnetic field, that is the transformation $I$ of \cite{enhanced}, where the complex parameter is $\l=\frac{b^2-4i\jmath}{b^2+4i\jmath}$.} is, up to a additive gauge constant,
\beq \label{A-swirling-melvin}
        A_\m dx^\m = \frac{8b[4b^2+(b^4+16\jmath^2)\r^2](4\jmath z d\tilde{t} - \D_\phi d\tilde{\phi})}{(b^4+16\jmath^2)(16+8b^2\r^2+(b^4+16\jmath^2)\r^4)} \ .
\eeq
A vast group of black holes can be embedded in this background, endowed with electromagnetic charges, acceleration \cite{illy} and angular momentum \cite{adriano-andrea}.\\

\paragraph{Bonnor-Melvin, k = 0 } When the swirling parameter $\jmath$ becomes zero, in the solution (\ref{g-swirling-melvin})-(\ref{A-swirling-melvin}), we have the Bonnor-Melvin universe \cite{bonnor}-\cite{melvin}
\beq \label{g-melvin}
          ds^2 = \left(1+\frac{b^2\r^2}{4} \right)^2 \left( - d\tilde{t}^2 + dz^2 + d\r^2 \right) + \frac{\r^2 d\tilde{\phi}^2}{\left(1+\frac{b^2\r^2}{4} \right)^2} ,
\eeq
with
\beq  \label{A-melvin}
            A = \left( A_{t_0}  - \frac{2b\r^2}{4+b^2\r^2} \right) d\tilde{\phi} \ .
\eeq
To get the above solution (\ref{g-melvin})-(\ref{A-melvin}) from double Wick rotation of the topological Kerr-Newman-NUT (\ref{KN-dwr})-(\ref{A-dwr}) we have to vanish the cosmological constant, make the following change of coordinates and adjustment of the integrating constants\footnote{This choice of the parametrisation is not unique, for instance it is possible to get the Bonnor-Melvin spacetime from the parametrisation used to get the Swirling Bonnor-Melvin universe with $\jmath=0$.}
\bea
           r &=& 1 + \frac{b^2\r^2}{4}  \ , \hspace{1.5cm} x \ = \ z \ , \hspace{1.9cm} \phi \ = \ - \frac{2 \tilde{\phi}}{b}  \ , \hspace{1.5cm} \ell \ = \ a \ = \ 0 \ , \nn \\ 
           m  &=& - \frac{b^2}{2}  \ , \hspace{2.2cm}   \hat{e} \ = \ -b \ , \hspace{1.4cm} \D_\phi \ = \ - \frac{b^2}{2} \ , \hspace{1.5cm}  \ \hat{p} \ = \ \om_0 \ = \ 0 \ . \hspace{1.6cm} 
\eea
A large family of black holes in this background have been found, see for instance \cite{ernst-magnetic}, \cite{ernst-wild}, \cite{ernst-remove}, \cite{pair-rotating}.\\

\paragraph{Swirling universe, k = 0 } To reach the swirling universe metric 
\beq \label{swirling}
         ds^2 = \left(1 + \jmath^2 \r^4 \right) \left( - d\tilde{t}^2 + dz^2 + d\r^2 \right) + \frac{\r^2(\D_\phi d\tilde{\phi} -4\jmath z d\tilde{t})^2}{1 + \jmath^2 \r^4} \ ,
\eeq
from  (\ref{KN-dwr})-(\ref{A-dwr}) with $\L=0$, the following change of coordinates can be used
\bea
        t &=& \sqrt{\jmath} \ \tilde{t} \ , \hspace{1.5cm} r \ = \ - \sqrt{\jmath} \ \r^2 \ , \hspace{1.5cm} x \ = \ - \sqrt{\jmath} \ z \ , \hspace{1.5cm} \phi \ = \ - \frac{\hat{\phi}}{2\sqrt{\jmath}} \ , \nn \\
               m &=& \frac{2}{\sqrt{\jmath}} \ , \hspace{1.7cm} \ell \ = \ \frac{1}{\sqrt{\jmath}}  \ , \hspace{2.1cm} \hat{e} \ = \ \hat{p} \ = \ 0 \ , \hspace{1.5cm} a \ = \ \om_0 \ = \ 0 \ .
\eea
The accelerating Kerr-Newman black hole class, with all its subcases has been embedded into the swirling universe \cite{swirling}, \cite{removal} and even a rotating binary black hole system at equilibrium \cite{swirling-binary}. \\

\paragraph{Minkowski}

The Minkowski spacetime can be obtained in several ways, for instance from the above Bonnor-Melvin universe in eq (\ref{g-melvin}) switching off the external electromagnetic field, by setting $b=0$, or by vanishing the swirling parameter $\jmath$ into the eq (\ref{swirling}). \\

\paragraph{Levi-Civita, k = 0 } The $\s=1$ Levi-Civita \cite{stephani-big-book} metric
\beq \label{levi-civita-rho}
             ds^2 =  - \r^{4\s} d \tilde{t}^2 + \r^{4\s(2\s-1)} (d\r^2+dz^2) + \r^{2(1-2\s)} d\tilde{\phi^2}
\eeq
can be obtained  from (\ref{KN-dwr})-(\ref{A-dwr}) with the following changes
\bea
        r &=& - \r^2  \ , \hspace{1.5cm} x \ = \ -z \ , \hspace{2.2cm} \phi \ = \ -\frac{\tilde{\phi}}{2} \ , \hspace{1.5cm} m \ = \ 2 \ , \\
        \ell &=& 0 \ , \hspace{2cm} \hat{e} \ = \ \hat{p} \ = \ 0 \ , \hspace{1.6cm} a \ = \ 0 \ , \hspace{1.8cm} \om_0 \ = \ 0 \ .
\eea
\\

\paragraph{Bertotti-Robinson-Bonnor-Melvin, $\bf k = -B^2$ }
When, in the general background (\ref{KN-dwr})-(\ref{A-dwr}), we set $\hat{p}= a = \om_0 = \L = 0$, we get the Levi-Civita-Bertotti-Robinson-Bonnor-Melvin spacetime, which is a superposition of the Levi-Civita-Bertotti-Robinson\footnote{Even though Levi-Civita was the first to discover the conformally flat spacetime with an homogeneous electromagnetic field in general relativity \cite{levi-civita-BR}, often in the literature, his name is dropped and this solution is known after Bertotti and Robison.} and the Bonnor-Melvin universe, recently built in \cite{bh+BRBM}. To write it as in \cite{bh+BRBM}, 
\beq \label{BRBM-metric}
 ds^2 = \Xi^2 \left[-(1+B^2\tilde{r}^2) dt^2 + \frac{d\tilde{r}^2}{1+B^2\tilde{r}^2} +\frac{\tilde{r}^2 d\tilde{x}^2}{(1-\tilde{x}^2)}\right] + \frac{\tilde{r}^2(1-\tilde{x}^2)}{\Xi^2 \ \big[1+B^2\tilde{r}^2(1-\tilde{x}^2)\big]^2} d\tilde{\phi}^2 \ ,
\eeq
with
\beq
           \Xi(\tilde{r},\tilde{x}) = \frac{-b(b+2B)+(b^2+2bB+2B^2)\sqrt{1+B^2\tilde{r}^2(1-\tilde{x}^2)}}{2B^2[1+B^2\tilde{r}^2(1-\tilde{x}^2)]} \ ,
\eeq
\beq \label{BRBM-A}
           A_\m = \left\{ 0,\, 0, \, 0,\, \frac{2(b+B)\tilde{r}^2(1-\tilde{x}^2)}{-(b^2+2bB+2B^2) \tilde{r}^2 (1-\tilde{x}^2)+2\Big[1+\sqrt{1+B^2\tilde{r}^2(1-\tilde{x}^2)}\Big]}  \right\} \ .
\eeq
From (\ref{KN-dwr})-(\ref{A-dwr}) with $\hat{p}= a = \om_0 = \L = 0$, one just needs to perform the following change of coordinates
\bea
          r &=& 1 - \frac{b(b+2B)\left[1-\sqrt{1+B^2\tilde{r}^2(1-\tilde{x}^2)} \right]}{2B^2\sqrt{1+B^2\tilde{r}^2(1-\tilde{x}^2)}} \ , \\
          x &=& \frac{\tilde{r}\tilde{x}}{\sqrt{1+B^2\tilde{r}^2(1-\tilde{x}^2)}} \ , \\
     \phi &=& \frac{2}{b^2+2bB} \, \tilde{\phi}  \ ,
\eea 
and redefinition of the parameters
\beq
       m = - \frac{b^2}{2} \ , \hspace{1.5cm} k = -B^2 \ , \hspace{1.5cm} k_7=B^2+B^4 \ , \hspace{1.5cm}
       \hat{e} = (b+B) \ .
\eeq
Note that from the Bertotti-Robinson-Bonnor-Melvin spacetime (\ref{BRBM-metric})-(\ref{BRBM-A}), when the parameter $B=0$ we obtain again the Bonnor-Melvin universe
\beq \label{BM-g}
         ds^2 = \left[1+\frac{b^2 \tilde{r}^2}{4} (1-\tilde{x}^2)\right]^2 \left[- dt^2 + d\tilde{r}^2 +\frac{\tilde{r}^2 d\tilde{x}^2}{(1-\tilde{x}^2)}\right] + \frac{\tilde{r}^2(1-\tilde{x}^2)}{\left[1+\frac{b^2 \tilde{r}^2}{4} (1-\tilde{x}^2)\right]^2} d\tilde{\phi}^2 \ ,
\eeq
\beq \label{BM-A}
           A_\m = \left[ 0,0,0, \frac{2b\tilde{r}^2(1-\tilde{x}^2)}{4+b^2\tilde{r}^2(1-\tilde{x}^2)} \right] \ .
\eeq
On the other hand when $B\neq 0$ and $b=0$ (or also $b=-2B$) we recover the Bertotti-Robinson electromagnetic solution
\bea
          ds^2 &=& \frac{1}{1+B^2\tilde{r}^2(1-\tilde{x}^2)} \left[ -(1+B^2\tilde{r}^2)dt^2 + \frac{d\tilde{r}^2}{(1+B^2\tilde{r}^2)} + \frac{\tilde{r}^2 d\tilde{x}^2 }{(1-\tilde{x}^2)} + \tilde{r}^2(1-\tilde{x}^2) \ d\tilde{\phi}^2 \right] \ , \\
         A_\m &=& \left[ 0, 0, 0, - \frac{B\tilde{r}^2(1-\tilde{x}^2)}{1+B^2\tilde{r}^2(1-\tilde{x}^2)+\sqrt{1+B^2\tilde{r}^2(1-\tilde{x}^2)}}  \right] \ .
\eea
Hence it is clear that the parameters $b$ and $B$ represents the intensity of the Bonnor-Melvin and Bertotti-Robinson electromagnetic fields respectively. Several black holes in the Bertotti-Robinson background has been discovered, see \cite{kerr-bertotti}, \cite{Ovcharenko:2025cpm}, \cite{garcia-alekseev}, \cite{alekseev-RN-bertotti}. To obtain the swirling generalisation of the Levi-Civita-Bertotti-Robinson-Bonnor-Melvin background one needs just to leave the $\ell\neq0$. Black holes in these backgrounds have been also generated in \cite{bh+BRBM}.\\

\paragraph{A hyperbolic pure gravitational backgrounds, $\bf k = -B^2$}

At the special point where $b=-B$, the Levi-Civita-Bertotti-Robinson-Bonnor-Melvin electrovacuum solution  (\ref{BRBM-metric})-(\ref{BRBM-A}) has a vanishing electromagnetic field \cite{bh+BRBM}, so it becomes just a vacuum metric representing a spacetime similar to the Witten bubble of nothing, but instead of being the double Wick rotation ($t\to i \phi, \, \phi \to i t$) of the standard Schwarzschild black hole with a spherical event horizon, it is a the double Wick rotation of the Schwarzschild metric with negative constant curvature in the section ($x,\phi$), as the negative value of $k$ denotes. The metric reads
\beq \label{background}
      ds^2 = \left[\frac{1+\sqrt{1+B^2\tilde{r}^2(1-\tilde{x}^2)}}{2\left( 1+B^2\tilde{r}^2(1-\tilde{x}^2 \right)} \right]^2 \left[-(1+B^2\tilde{r}^2) \, dt^2 + \frac{d\tilde{r}^2}{1+B^2\tilde{r}^2} + \frac{\tilde{r}^2 d\tilde{x}^2}{1-\tilde{x}^2} \right] + \frac{4\tilde{r}^2(1-\tilde{x}^2) \ d\varphi^2}{\left(1+\sqrt{1+B^2\tilde{r}^2 (1-\tilde{x}^2)}\right)^2}  \ .  
\eeq
The parameter $B$, in this case represents the intensity of the gravitational field, for $B=0$ the flat spacetime is recovered. When $B \to iB$ the sign of the curvature reverse and the metric describes a Witten's bubble of the next paragraph, for more details see to \cite{static-typeI-bh}.\\
Also in this case a static black hole can be embedded in this background \cite{static-typeI-bh}.\\

\paragraph{Expanding bubbles of nothing, k = 1 }

When $k=1$ we get a generalisation of the expanding bubble of nothing \cite{witten}, \cite{horowitz-bubble-baths}. These extensions can include the
NUT parameter, the angular momentum, the charge. The simpler case is the original Witten one that can be obtained, from  (\ref{KN-dwr})-(\ref{A-dwr}), when all the charges are set to zero, apart from the mass: $m=r_0/2, \ a=\ell=\hat{e}=\hat{p}=0$. In this case, thanks to the change of coordinates
\beq
         t = \textrm{arccsch} \left[ \cos \chi \ \textrm{csch} T \sqrt{1-\sinh^2 T \tan^2 \chi} \right] \ , \hspace{1.6cm}  x= \cosh T \, \sin \chi
\eeq
we remain with the standard form of the Witten bubble of radius $r_0$
\beq
       ds^2 =  r^2\left( - dT^2  + \cosh^2 T \, d\chi^2 \right) +  \frac{dr^2}{1-\frac{r_0}{r}}+\left(1-\frac{r_0}{r} \right) d\phi^2 \ .
\eeq
Static, rotating and multi-black holes at equilibrium in this background can be built, see \cite{bubble} and \cite{kerr+bubble}.\\

\subsection{A novel rotating background for $k=0$: Bonnor-Melvin swirling curling universe}
\label{bonnor-melvin-swirling-curling}

In all the spacetimes considered above there is a physical parameter which has been always kept null: $a$, the reminiscent of the angular momentum of the Kerr black hole, before the analytic continuation. It might be interesting to explore the possibility of leaving this rotational parameter, which is independent from the swirling one, switched on, in order to model novel kinds of stationary gravitational backgrounds. This is the aim of this section.\\
In particular we present here a, three parameters, specialization of the general background which includes the Bonnor-Melvin electromagnetic field and a couple of independent rotational external backgrounds: the swirling one, labelled with the character $\jmath$, and the new one, which we call $curling$, denoted by the symbol $\aleph$. We write here explicitly the case for  $k=0$, which is the direct generalisation of the swirling Bonnor-Melvin universe, for more general cases $k$ can be chosen differently. For this purpose the following coordinates transformation and redefinitions of parameters, from (\ref{KN-dwr})-(\ref{A-dwr}), is needed:
\bea
        r &=& \frac{(b^4+16\jmath^2)^{1/4}}{8} \left(\aleph^2-4\r^2 -\frac{16b^2}{b^4+16\jmath^2} \right) \ , \hspace{2cm} x \ = \ - \frac{z}{2} (b^4+16\jmath^2)^{1/4}  \ ,\nn \\
         \phi &=& \frac{-\D_\phi \tilde{\phi}}{(b^4+16\jmath^2)^{1/4}} \ ,        \hspace{5.9cm} t \  = \ \frac{\tilde{t}}{2} (b^4+16\jmath^2)^{1/4}  \ ,  \nn \\ 
          \hat{e} &=& \ \frac{4b}{(b^4+16\jmath^2)^{1/2}} ,  \hspace{5.9cm}  \ell \ = \ \frac{8\jmath}{(b^4+16\jmath^2)^{3/4}}  \ ,\nn \\
            m &=& \ \frac{4}{(b^4+16\jmath^2)^{1/4}}   \ , \hspace{5.8cm} \hat{p} \ = \ \om_0 \ = \ 0 \ ,  \nn \\
     a &=& - \ \aleph  \ .
\eea
The resulting metric is 
\beq
       ds^2 = \frac{\r^2\D_\phi^2 d\tilde{t}^2}{f(\r,z)} +   \frac{\D_\phi^2(\aleph^2-4\r^2)}{4f(\r,z)}  \Big( d\r^2 +dz^2 \Big)
- f(\r,z)  \left\{ \left[ \frac{\D_\phi \aleph}{2 f(\r,z)} -\frac{z[b^4z\aleph + 16\jmath(2+\jmath z \aleph)]}{8\D_\phi} \right]d\tilde{t} + d\tilde{\phi} \right\}^2  \ , 
\eeq
with
\beq
          f(\r,z) \ = \ \frac{64\D_\phi^2(\aleph^2-4\r^2)}{256 + 32b^2(4\r^2-\aleph^2) + b^4 [16z^2\aleph^2+(\aleph^2-4\r^2)^2]+16\jmath\{32z\aleph+\jmath[16z^2\aleph^2+(\aleph^2-4\r^2)^2]\}} \ .
\eeq
Clearly, when the curling parameter $\aleph=0$, we recover the swirling Bonnor-Melvin background (\ref{g-swirling-melvin})-(\ref{A-swirling-melvin}). \\ 
However note that the chosen parametrisation may be improved because, while the limits to the swirling Bonnor-Melvin background are straightforward and well defined, for $\aleph=0$, the curling Bonnor-Melvin and the curling swirling backgrounds go directly to the Minkowski spacetime for vanishing $b$ and $\jmath$, more details about that in the section \ref{sec:bh+bacground}.  \\
  
Note that all the above background subcases can be straightforwardly extended in the presence of the cosmological constant, just leaving non-null $\L$ form the general background  (\ref{KN-dwr})-(\ref{A-dwr}), as it has been noticed for the Bonnor-Melvin and the swirling cases in \cite{charging} and \cite{swirling} respectively. However in the presence of the cosmological constant the superposition of a black hole with any of these backgrounds is difficult because some symmetries of the gravitational field equations, fundamental for developing useful solution generating techniques, are broken \cite{charging}. \\

\section{Embedding Schwarzschild in the novel rotating background}
\label{sec:bh+bacground}

\subsection{Schwarzschild in the curling, Bonnor-Melvin universe}

Having unified and clarified the relation between the black hole backgrounds in general relativity, it would be interesting to embed inside these spacetimes a black hole metric. We will insert the Schwarzschild black hole into the above background solution with $\jmath=0$, just to keep the computations simple. Nevertheless there are no conceptual or computational obstacles in considering charged and rotating black holes or keeping the background more general.\\ 
Thanks to the inverse scattering technique \cite{belinski-book} we can build the metric and the electromagnetic potential representing the superposition of the Schwarzschild black hole and the Bonnor-Melvin electromagnetic background, which, in spherical coordinates, read\footnote{A Mathematica notebook containing this solution is available between the arXiv source files.}
\beq \label{curlin-melvin-schwarzschild-inizio}
  ds^2 =  -f(r,x)\big[d\varphi -\omega(r,x)dt \big]^2 +  \frac{1}{f(r,x)} \left\{ \r^2(r,x) dt^2 - e^{2\g(r,x)} \left[ \frac{dr^2}{\Delta_r(r)} + \frac{dx^2}{\Delta_x(x)} \right] \right\} 
\eeq  
\bea
      f(r,x) &=&  - \left[256 r^4 (1-x^2) - 64 r (r-2m) \aleph^2 \right]/D(r,x) \nn \\
      \omega(r,x)&=& \frac{b^4 \aleph}{32} \left\{ r(r-6m)(1-x^2)-12m^2(1+x^2) + \aleph^2 + \frac{16 \big[8m-4r+b^4r^3(r-3m)^2(x^4-x^2)\big]}{b^4 \big[ (r-2m)\aleph^2 - 4r^3(1-x^2) \big]} \right\} + \om_0 \nn \\
     \gamma(r,x)  &=&  \frac{1}{2} \log \left\{ \frac{r}{4} \Big[4r^3(1-x^2)-(r-2m)\aleph^2 \Big] \right\} + \gamma_0 \ , \label{gamma} \\
     \r(r,x) &=&  \sqrt{\D_r(r) \D_x(x)} \ , \hspace{2.2cm} \D_r(r) = r^2 -2mr \ , \hspace{2.2 cm} \D_x(x) = 1-x^2 \ , \nn \\
     D(r,x) &=& 16 b^4 r^2 (r-3m)^2 x^2 \aleph^2 + \big[b^2  \aleph^2 (r-2m) - 4 r (4 + b^2 r^2 (1-x^2)) \big]^2 \ , \nn  \\
     A_t(r,x) &=& \bigg\{32br[4+b^2r^2(1-x^2)]^2\left\{4r+2b^2r^3 + 6 b^2 m^2 r (2x^2-1) + m [4-3b^2r^2(1+3x^2)]   \right\} \aleph     \nn \\
     &+& 4b^3\left\{ 4r^2+3b^2r^4(1-5x^2) +24b^2m^3r(1+x^2) -4m^2(4+33b^2r^2x^2)\right\} \aleph^3 \nn  \\
     &-& 16b^3mr[8-b^2r^2(5-23x^2)] \aleph^3 + b^5(r-2m)(6m+r)\aleph^5 \bigg\} /[4D(r,x)] + A_{t0}  \ , \nn \\ 
     A_\varphi(r,x) &=& \left\{32r\big[16r + 4b^2r^3(1-x^2) - b^2 (r-2m) \aleph^2 \big] \right\}/ [b D(r,x)] + A_{\varphi0} \ . \label{curlin-melvin-schwarzschild-fine}
\eea
We recall that $A_{t0}, \ A_{\varphi0}$, $\g_0$ and $\om_0$ are just gauge constants\footnote{In general these quantities are can be thought null, unless elsewhere specified.} (always present in general relativistic metrics written in Lewis-Weyl-Papapetrou form) which can be useful to regularise a possible conical singularity or to set the angular velocity of an observer at spatial infinity, respectively.\\
Since the metric contains the Schwarzschild-Bonnor-Melvin, it must be of general type I.\\
Possible divergences of the Kretschmann scalar invariant, for the above solution, are located where its denominator, 
\beq
 \frac{16 b^4 r^2 x^2 (r-3 m)^2}{\aleph ^2}+\left\{b^2 (r-2 m)-\frac{4 r\left[b^2 r^2 \left(1-x^2\right)+4\right]}{\aleph ^2}\right\}^2 \ , \nn
\eeq
is null. Since it is the sum of two squares, it can be zero only if both the two squares are simultaneously vanishing. This can not occur for a significant portion of the parametric space. Therefore the background geometry, in particular the novel features coming from the presence of $\aleph$ can smear the typical curvature singularity in $r=0$ of the Schwarzschild black hole, in a certain sector of the parametric space. This is not a complete novelty, because it can occur also in the context of Bonnor-Melvin and swirling backgrounds \cite{adriano-andrea}. \\
 
The $\aleph \to 0$ limit (adjusting properly the gauge constants) of the above solution recovers the Schwarzschild black hole embedded into the magnetic Bonnor-Melvin universe
\beq \label{melvin-bh-ds}
          ds^2 = \left[1+\frac{b^2 r^2}{4} (1-x^2)\right]^2 \left[- \left(1- \frac{2m}{r} \right) dt^2 + \frac{e^{2\g_0}dr^2}{1-\frac{2m}{r}} +\frac{e^{2\g_0} r^2 dx^2}{(1-x^2)}\right] + \frac{r^2(1-x^2) \, d\varphi^2}{\left[1+\frac{b^2 r^2}{4} (1-x^2)\right]^2}  \ ,
\eeq 
\beq\label{melvin-bh-A}
           A_\m = \left[ 0,0,0, \frac{2br^2(1-x^2)}{4+b^2r^2(1-x^2) } + A_{\varphi0}\right] \ .
\eeq
In this $\aleph=0$ subcase the usual curvature singularity of the static black hole is also retrieved. Also the electromagnetic background is completely regular. Clearly the above solution (\ref{melvin-bh-ds})-(\ref{melvin-bh-A}) reduces to the magnetic Bonnor-Melvin universe when we further switch off the black hole mass $m$, as written in eqs. (\ref{BM-g})-(\ref{BM-A}). 
On the other hand, the limit for $b \to 0$ gives the Schwarzschild black hole (in a rotating frame of reference; to have it in the usual static gauge, an extra gauge constant has to be set to zero). This fact is independent of whether  we perform the limit starting from (\ref{melvin-bh-ds})-(\ref{melvin-bh-A}) or from  (\ref{curlin-melvin-schwarzschild-inizio})-(\ref{curlin-melvin-schwarzschild-fine}), because at this level the parameters $b$ and $\aleph$ are somehow intertwined. Obviously when the black hole is removed ($m=0$) from the spacetime (\ref{curlin-melvin-schwarzschild-inizio})-(\ref{curlin-melvin-schwarzschild-fine}) we have the Bonnor-Melvin curling background, which is a non-swirling subcase of the spacetime of section \ref{bonnor-melvin-swirling-curling}. Note however that without the swirling contribution this background can display curvature singularities on the equatorial plane, for $\r=\frac{\sqrt{b^2\aleph^2-16}}{2b}$. Anyway, for $b^2\aleph^2<16$, there are no divergences in the curvature scalar invariant. From a physical perspective it means that the model is well behaved in a regime where the background rotation is bounded.  \\
So, even though the solution (\ref{curlin-melvin-schwarzschild-inizio})-(\ref{curlin-melvin-schwarzschild-fine}) clearly contains the Schwarzschild metric, the black hole interpretation for this spacetime is not guaranteed, for all the range of the physical parameters. \\

\subsection{Schwarzschild embedded in a curling background}
\label{schwarzschild-kerrling}

The removal of the Bonnor-Melvin background is not so direct, in this parametrization, because of the above mentioned entanglement between $b$ and $\aleph$. However it can be done in two different approaches, which give an equivalent result: by applying the Harrison transformation to solution (\ref{curlin-melvin-schwarzschild-inizio})-(\ref{curlin-melvin-schwarzschild-fine}) or by taking its limit for $b\to\infty$\footnote{To select the right leading order in the limiting procedure we rescale properly the Killing coordinates and the gauge constants. More specifically $\om_0 \to \om_0 b^4, \ t \to 4t/b^2, \ \varphi \to \varphi b^2/4, \ \g_0=2\log2/b)$}. In any case the result represents a Schwarzschild metric embedded only in the novel rotating background, which was the last remained  unexplored background related to the Kerr-Newman metric with a flat base angular manifold\footnote{We recall that the background related to the conjugate solution with respect to the flat (extremal) Reissner-Nordstrom is the Bonnor-Melvin background \cite{bonnor}-\cite{melvin} and the swirling background of \cite{harrison}. The black hole solutions in these backgrounds were discovered in \cite{ernst-magnetic} and \cite{swirling} respectively.}, the one related to the angular momentum parameter $a$. Notice that the electromagnetic field completely vanishes in this limit. It can be written\footnote{A Mathematica notebook containing this solution is available between the arXiv source files.} from the general metric (\ref{curlin-melvin-schwarzschild-inizio}),(\ref{gamma}) and
\bea \label{sch-kerling-f}
          f(r,x) &=& \frac{- 16 r^4 (1-x^2) + 4 r (r - 2 m) \aleph^2 }{ 16 r^6 (1-x^2)^2 + 8 r^2 \big[ r(2m-r) + (18m^2 -14mr + 3r^2)x^2 \big] \aleph^2 + (r-2m)^2 \aleph^4  } \ , \\
          \omega(r,x) &=& \left[ r \left(\frac{r}{2} -3m \right) (1-x^2) -6m^2(1+x^2) \right] \aleph + \frac{8r^3(r-3m)^2 x^2 (x^2-1) \aleph}{4r^3(x^2-1) + (r-2m) \aleph^2} + \om_0  \ . \label{sch-kerling-w}
\eea 
The Kretschmann scalar invariant for this metric is finite because for large radial distance it goes to zero, while for finite $r$ its denominator
\beq
           \frac{144 m^2 r^2 x^2}{\aleph ^2}+\frac{16 m r^3 \left(1-7 x^2\right)}{\aleph ^2}+(r-2 m)^2+\frac{16 r^6 \left(1-x^2\right)^2}{\aleph ^4}+\frac{8 r^4 \left(3 x^2-1\right)}{\aleph ^2}
\eeq
is positive for a large spectrum of the parametric space $r > 2 m > 0 $ and $m>\aleph/3$. Therefore, similarly to the previous section spacetime, the spacetime is not divergent, for a significant range of physical parameters. The effect of the external background is to remove curvature singularity, typical of the Schwarzschild black hole. \\

The above metric gives the pure $curling$ background when $m=0$. To write in Weyl coordinates ($\r,z$)
\beq\label{curling-1}
      ds^2 = -f (d\varphi - \om dt)^2 + \frac{1}{f} \left[-e^{2\g} (d\r^2 +dz^2) + \r^2 d\varphi^2 \right]
\eeq
where
\beq \label{curling-2}
      f = \frac{-4(4\r^2-\aleph^2)}{(4\r^2 -\aleph^2)^2 + 16z^2 \aleph^2} \ , \qquad \om = \frac{\r^2\aleph(16z^2+4\r^2-\aleph^2)}{8\r^2-2\aleph^2} \ , \qquad \g = \frac{1}{2} \log \left(\r^2-\frac{\aleph^2}{4} \right) \nn ,
\eeq
the following change of coordinates have been used
\beq \label{rho-z}
       r = \sqrt{\r^2+z^2} \ , \hspace{1.2cm} x = \frac{z}{\sqrt{\r^2+z^2}} \ .
\eeq
Note that the Weyl coordinates are not cylindrical for this metric, in order to put it in cylindrical coordinates we can further shift the $\r$ coordinate $\r \to \tilde{\r} = \sqrt{\bar{\r}^2+\aleph^2}/2 $.\\
The background, however, is less regular with respect to the $m\neq0$ spacetime, in fact scalar invariants diverge on the equatorial plane ($x=0$) for $\r=\aleph/2$. So at his respect the above section background, in the presence of the Bonnor-Melvin electromagnetic field, behaves better, for instance for small values of the magnetic field parameter $b$.\\
 
The spacetime (\ref{curlin-melvin-schwarzschild-inizio}),(\ref{sch-kerling-f})-(\ref{sch-kerling-w}) belongs to the general Petrov type, and its behaviour for large values of the coordinate $r$ is not depending on the mass parameter $m$, so it is shared by the curling background (\ref{curling-1})-(\ref{curling-2}). At the first leading order, it is given by\footnote{Of course this metric should not satisfy the Einstein equations, unless $\aleph=0$.}
\beq
       ds^2 =  (1-x^2)^2 r^4 \left[ - dt^2  + dr^2 + \frac{r^2}{1-x^2} \right] + \frac{d\varphi^2}{r^2(1-x^2)} - \frac{1+3x^2}{1-x^2} \aleph dtd\varphi \ .
\eeq
In terms of the coordinated defined in (\ref{rho-z}) the above metric can be written as 
\beq 
       ds^2 = \r^4 \left( -dt^2 + d\r^2 + dz^2 \right) + \r^{-2} d\varphi^2 - \aleph \left(1 + 4z^2\r^{-2} \right) dtd\varphi \  \nn .
\eeq
Thus it resembles a rotating Levi-Civita background.\\
The Schwarzschild black hole 
\beq
               ds^2 = - \left( 1 - \frac{2m}{\bar{r}} \right) dt^2 + \frac{dr^2}{1-\frac{2m}{\bar{r}}} + \bar{r}^2 d\theta^2 + \bar{r}^2 \sin^2\theta d\varphi^2
\eeq
can be retrieved from (\ref{curlin-melvin-schwarzschild-inizio}),(\ref{sch-kerling-f})-(\ref{sch-kerling-w}) for $\aleph=\infty$\footnote{Obviously defining a new parameter related to the rotation as $\Xi = 1/\aleph$ for the Schwarzschild curling metric  (\ref{curlin-melvin-schwarzschild-inizio}),(\ref{sch-kerling-f})-(\ref{sch-kerling-w}) would allow one to recover the static, spherically symmetric black hole in the more intuitive notation $\Xi \to 0$.}. Before taking that limit, to get a clearer result, the following rescaling of the Killing coordinates, shift of the radial coordinate and properly fixing of the gauge parameters have been performed 
\beq
         t = \frac{2}{\aleph} t \ , \qquad \quad \varphi = \frac{\aleph}{2} \varphi \ , \qquad \quad \om_0 = \frac{3m^2\aleph}{8} - \frac{\aleph^3}{32} \ , \qquad \quad \g_0 = \frac{1}{2} \log\left( \frac{16}{\aleph^4} \right) \ , \qquad \quad r = 2m - \bar{r} \ .         
\eeq
On the other  $\aleph \to 0$ limit gives the Schwarzschild solution into the Levi-Civita background \cite{maccallum}
\beq \label{sch-LC-ds}
           ds^2 = r^4(1-x^2)^2 \left[ - \left(1 -\frac{2m}{r} \right) dt^2 + \frac{dr^2}{1-\frac{2m}{r}} + \frac{r^2 dx^2}{1-x^2}  \right] + \frac{d\varphi^2}{r^2(1-x^2)} \ .
\eeq 
Note however that this metric is not suitable for black hole interpretation because it displays naked, unavoidable curvature singularities on the axis of symmetry ($x=\pm1$). Actually it represents a physically pathological limit of the Schwarzschild swirling black hole solution, which contains it \cite{bh+BRBM}. We recall that the Schwarzschild swirling does not present curvatures or conical singularities outside the event horizon.  \\
More generally, black holes into the Levi-Civita background can be retrieved from any black hole of this class, that is black holes in Bonnor-Melvin, swirling or curling background, for more details about that see the appendix \ref{app:levi-bh}. This is fundamentally due to the common asymptotic behaviour this class of background possesses. \\

\section{Including acceleration}

\subsection{Self duality of the C-metric}

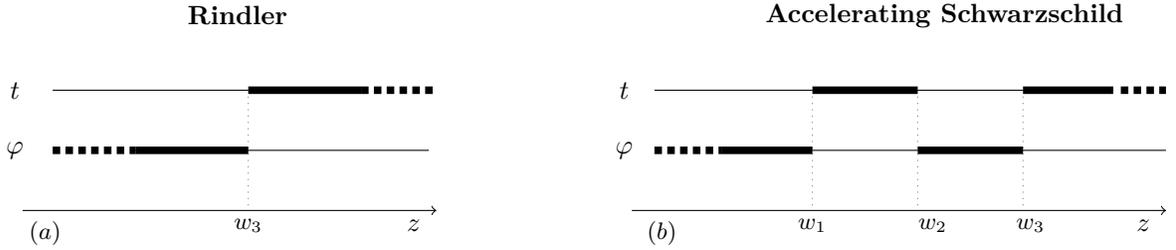
\begin{figure}[h!]
\centering
\begin{tikzpicture}
 
% da papero equivalnece

\draw[black,thin] (-10.1,-2) -- (-6.1,-2);
\draw[black,thin] (-9,-2.8) -- (-5.1,-2.8);
\draw[black,->] (-10.5,-3.6) -- (-5,-3.6);

\draw (-10.6,-2) node{$t$};
\draw (-10.6,-2.8) node{$\varphi$};
\draw (-5.3,-3.8) node{$z$};
\draw (-10.2,-3.9) node{{\small $(a)$}};
\draw (-7.5,-3.8) node{{\small $w_3$}};
%\draw (-8,-3.8) node{{\small $w_2$}};
%\draw (-7,-3.8) node{{\small $w_3$}};
%\draw (-6,-3.8) node{{\small $w_4$}};
\draw (-7.7,-1) node{{{ \bf Rindler}}};

\draw[gray,dotted] (-7.5,-2) -- (-7.5,-3.6);
%\draw[gray,dotted] (-8,-2) -- (-8,-3.6);
%\draw[gray,dotted] (-7,-2) -- (-7,-3.6);
%\draw[gray,dotted] (-6,-2) -- (-6,-3.6);

\draw[black, dotted, line width=1mm] (-10.1,-2.8) -- (-9,-2.8);
\draw[black,line width=1mm] (-9,-2.8) -- (-7.5,-2.8);
\draw[black,line width=1mm] (-7.5,-2) -- (-6,-2);
\draw[black, dotted,line width=1mm] (-6,-2) -- (-5,-2);
%\draw[black,line width=1mm] (-8,-2.8) -- (-7,-2.8);
%\draw[black,line width=1mm] (-7,-2) -- (-6,-2);
%\draw[black,line width=1mm] (-6,-2.8) -- (-5,-2.8);&
%\draw[black, dotted,line width=1mm] (-5,-2.8) -- (-4.4,-2.8);

%\draw[black,->] (-3.5,-2.8) -- (-1.7,-2.8) node[midway, below, sloped] {{\small $w_4 \to +\infty$}};

\draw[black,thin] (-2.1,-2) -- (4,-2);
\draw[black,thin] (-1.2,-2.8) -- (4.7,-2.8);
\draw[black,->] (-2.4,-3.6) -- (4.7,-3.6);

\draw (-2.5,-2) node{$t$};
\draw (-2.5,-2.8) node{$\varphi$};
\draw (4.4,-3.8) node{$z$};

\draw[gray,dotted] (0,-2) -- (0,-3.6);
\draw[gray,dotted] (1.4,-2) -- (1.4,-3.6);
\draw[gray,dotted] (2.8,-2) -- (2.8,-3.6);

\draw (-2,-3.9) node{{\small $(b)$}};
\draw (0,-3.8) node{{\small $w_1$}};
\draw (1.6,-3.8) node{{\small $w_2$}};
\draw (2.9,-3.8) node{{\small $w_3$}};

\draw[black, dotted, line width=1mm] (-2.1,-2.8) -- (-1.2,-2.8);
\draw[black,line width=1mm] (-1.2,-2.8) -- (0,-2.8);
\draw[black,line width=1mm] (0,-2) -- (1.4,-2);
\draw[black,line width=1mm] (1.4,-2.8) -- (2.8,-2.8);
\draw[black,line width=1mm] (2.8,-2) -- (4,-2);
\draw[black, dotted,line width=1mm] (4.1,-2) -- (4.7,-2);

\draw (1.7,-1) node{{ {\bf Accelerating Schwarzschild}}};

\end{tikzpicture}
\caption{{\small Rods diagram representing the conjugate Rindler metric $(\ref{fig:rod-c}.a)$ and the conjugate accelerating Schwarzschild black hole or the uncharged c-metric $(\ref{fig:rod-c}.b)$. The finite time-like segment represents the event horizon, while the accelerating horizon is represented by a infinite time-like semi-line.}}
\label{fig:rod-c}
\end{figure}

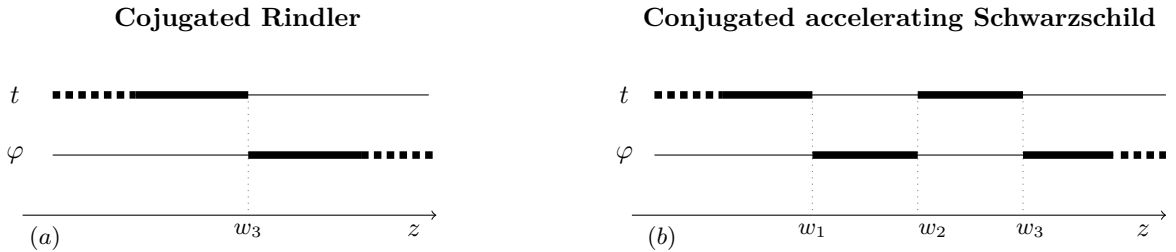
\begin{figure}[h!]
\centering
\begin{tikzpicture}
 
% da papero equivalnece

\draw[black,thin] (-10.1,-2.8) -- (-6.1,-2.8);
\draw[black,thin] (-9,-2) -- (-5.1,-2);
\draw[black,->] (-10.5,-3.6) -- (-5,-3.6);

\draw (-10.6,-2) node{$t$};
\draw (-10.6,-2.8) node{$\varphi$};
\draw (-5.3,-3.8) node{$z$};
\draw (-10.2,-3.9) node{{\small $(a)$}};
\draw (-7.5,-3.8) node{{\small $w_3$}};
%\draw (-8,-3.8) node{{\small $w_2$}};
%\draw (-7,-3.8) node{{\small $w_3$}};
%\draw (-6,-3.8) node{{\small $w_4$}};
\draw (-7.7,-1) node{{{ \bf Cojugated Rindler}}};

\draw[gray,dotted] (-7.5,-2) -- (-7.5,-3.6);
%\draw[gray,dotted] (-8,-2) -- (-8,-3.6);
%\draw[gray,dotted] (-7,-2) -- (-7,-3.6);
%\draw[gray,dotted] (-6,-2) -- (-6,-3.6);

\draw[black, dotted, line width=1mm] (-10.1,-2) -- (-9,-2);
\draw[black,line width=1mm] (-9,-2) -- (-7.5,-2);
\draw[black,line width=1mm] (-7.5,-2.8) -- (-6,-2.8);
\draw[black, dotted,line width=1mm] (-6,-2.8) -- (-5,-2.8);
%\draw[black,line width=1mm] (-8,-2.8) -- (-7,-2.8);
%\draw[black,line width=1mm] (-7,-2) -- (-6,-2);
%\draw[black,line width=1mm] (-6,-2.8) -- (-5,-2.8);&
%\draw[black, dotted,line width=1mm] (-5,-2.8) -- (-4.4,-2.8);

%\draw[black,->] (-3.5,-2.8) -- (-1.7,-2.8) node[midway, below, sloped] {{\small $w_4 \to +\infty$}};

\draw[black,thin] (-2.1,-2.8) -- (4,-2.8);
\draw[black,thin] (-1.2,-2) -- (4.7,-2);
\draw[black,->] (-2.4,-3.6) -- (4.7,-3.6);

\draw (-2.5,-2) node{$t$};
\draw (-2.5,-2.8) node{$\varphi$};
\draw (4.4,-3.8) node{$z$};

\draw[gray,dotted] (0,-2) -- (0,-3.6);
\draw[gray,dotted] (1.4,-2) -- (1.4,-3.6);
\draw[gray,dotted] (2.8,-2) -- (2.8,-3.6);

\draw (-2,-3.9) node{{\small $(b)$}};
\draw (0,-3.8) node{{\small $w_1$}};
\draw (1.6,-3.8) node{{\small $w_2$}};
\draw (2.9,-3.8) node{{\small $w_3$}};

\draw[black, dotted, line width=1mm] (-2.1,-2) -- (-1.2,-2);
\draw[black,line width=1mm] (-1.2,-2) -- (0,-2);
\draw[black,line width=1mm] (0,-2.8) -- (1.4,-2.8);
\draw[black,line width=1mm] (1.4,-2) -- (2.8,-2);
\draw[black,line width=1mm] (2.8,-2.8) -- (4,-2.8);
\draw[black, dotted,line width=1mm] (4.1,-2.8) -- (4.7,-2.8);

\draw (1.1,-1) node{{ {\bf Conjugated accelerating Schwarzschild}}};

\end{tikzpicture}
\caption{{\small Rods diagram representing the Rindler metric $(\ref{fig:dwr-rod-c}.a)$ and the accelerating Schwarzschild black hole or the uncharged c-metric.  The diagrams coincide with the non-conjugated ones of figure \ref{fig:rod-c} just by inverting the $z$ coordinate.}}
\label{fig:dwr-rod-c}
\end{figure}

The only parameter we have not explicitly considered yet is the acceleration. That's because it behaves differently with respect to the other parameters of the black hole solution. In fact, while the external properties of the backgrounds (\ref{KN-dwr}) are related with the associated double Wick rotated black holes charges, accelerating black holes (as also the Rindler background) are self dual under the conjugation transformation of the metric. This is quite obvious from the rod diagram representation of the metrics. For instance, if we write the rod representation of the accelerating black hole in Weyl coordinates $(\r,z)$ \cite{emparan-reall}, as in figure \ref{fig:rod-c} and their conjugate metrics as in figure \ref{fig:dwr-rod-c}, we can clearly see how the two diagrams are identical upon the inversion of the $z$ axis.

In fact the coordinate transformation $z \to - z$ is exactly the diffeomorphism which maps the Rindler metric
\beq  \label{rindler-c}
                           ds^2 =  - \m_3 \, dt^3  \ + \  \frac{\m_3}{\r^2+\m_3^2} \, (d\r^2+dz^2) \ + \ \frac{\r^2}{\m_3} \, d\varphi^2      \nn 
\eeq
and the neutral C-metric
\beq \label{c-weyl}
           ds^2 = - \frac{\mu_1\mu_3}{\mu_2} \ dt^2 +  \frac{16\, C_f \, \mu_1^3 \mu_2^3 \mu_3^3  \ (d\r^2+dz^2)}{\mu_{12} \mu_{23} W_{13}^2 W_{11} W_{22} W_{33} } +  \frac{\rho^2 \, \mu_2}{\mu_1 \mu_3} \ d\varphi^2 \ . \nn
\eeq 

to their respective conjugate metrics
\beq
           ds^2 = - \frac{\r^2}{\m_3} \, dt^2    \ + \  \frac{\m_3}{\r^2+\m_3^2} \, (d\r^2+dz^2) \ + \ \m_3 \, d\varphi^3      \ , \nn
\eeq
\beq
             ds^2 =-\frac{\rho^2 \, \mu_2}{\mu_1 \mu_3} \ dt^2 +  \frac{16\, C_f \, \mu_1^3 \mu_2^3 \mu_3^3  \ (d\r^2+dz^2)}{\mu_{12} \mu_{23} W_{13}^2 W_{11} W_{22} W_{33} } +  \frac{\mu_1\mu_3}{\mu_2}  \ d\varphi^2 \ . \nn
\eeq
Here we are using the standard solitonic inverse scattering notation \cite{belinski-book}, \cite{marcoa-equivalence}, where the basic building blocks are
\beq
\label{mui}
\mu_i = w_i-z+\sqrt{\rho^2 + (z-w_i)^2} \ , \quad \qquad \mu_{ij} = (\mu_{i}-\mu_{j})^2  \ , \quad \qquad W_{ij} = \rho^2 + \mu_i\mu_j \ . \nn
\eeq
For completeness we recall the standard physical parametrization
\beq
         w_1 = -m \ ,  \hspace{1cm} w_2 = m \ , \hspace{1cm} w_3=\frac{1}{2A} \ , \hspace{1cm}  C_f = \frac{m^2}{A^3} \ , \nn
\eeq
that gives, thank to the change of coordinates
\beq
              \r(r,\theta) \, = \,  \frac{\sqrt{(r^2-2mr)(1-A^2 r^2)(1 + 2 m A x)(1-x^2)}}{(1 - A rx)^2} \  ,  \hspace{1.2cm}   z(r,\theta) \, = \,   \frac{(A r + x) [r-m(1- A rx)]}{(1 - A rx)^2}  \ , \nn
\eeq
the accelerating Schwarzschild black hole, also known as the uncharged C-metric, in spherical coordinates
\beq \label{c-metric}
      ds^2 =   \frac{\displaystyle \left(1-\frac{2m}{r} \right) (A^2r^2-1) dt^2 + \frac{dr^2}{ \left(1-\frac{2m}{r} \right) (1-A^2r^2)} + \frac{r^2 \ dx^2}{(1-2mA x)(1-x^2)} + r^2 (1-2mA x) (1-x^2) \ d\varphi^2}{(1 - A rx)^2} \nn
\eeq

The Rindler parametrization is clearly achieved, from the C-metric, by removing the black hole for $ w_1 \to w_2 $, or equivalently vanishing the black hole mass parameter $m \to 0$. \\
We have seen that in the presence of the acceleration, the conjugation operation maps accelerating black holes into accelerating black holes, thus conjugated metric of accelerating black holes are not always suitable to model gravitational backgrounds, in the same spirit of what we have done in section \ref{general-background}. \\
So to include the acceleration in this picture and enlarge this framework to all\footnote{We are thinking to black holes that are not singular outside theirs simply connected event horizon. However the general class of backgrounds, as in (\ref{KN-dwr}), are suitable to host multi black hole systems \cite{swirling-binary}, \cite{kerr+bubble}. Accelerating black holes can present conical singularities, but thanks to the interaction with Bonnor-Melvin or swirling backgrounds they can be regularised \cite{ernst-remove}, \cite{removal}.} known stationary and axisymmetric black hole in the theory of general relativity, possibly coupled with the Maxwell electromagnetism, we have two possibilities. The first is to embed into the background class in (\ref{KN-dwr}) accelerating black holes, while the second, as detailed in the following section, can be to extend the conjugated topological Kerr-Newman-NUT metric (\ref{KN-dwr}) to the accelerating parameter. A composition of these two accelerating effects, while it is possible, might be redundant.

\subsection{Accelerating backgrounds as the conjugated Plebanski-Demianski metric}

The accelerating generalisation of the Kerr-Newman black hole is still a type D space-time, according to the Petrov classification, belonging to the Plebanski-Demianski metric family. For describing black holes the best parametrization of the Plebanski-Demianski spacetime is given in \cite{marcoa-equivalence}, \cite{mostgeneral}, see also \cite{PD-revisited} for a relation with the previous parametrisations. Anyway for describing the background here we can use the parametrisation of \cite{new-look-PD}, because more similar to the section \ref{sec:backgrounds} notation, whose metric and potential are
\beq \label{PD-ds2}
          ds^2 =\frac{[\omega dt + (r^2 + \omega \hat{\om}_0) d\varphi]^2 P(x) - [dt + (-x^2 \omega + \hat{\om}_0) d\varphi]^2 Q(r)}{R^2(r, x) \Omega^2(r, x)} + \frac{[dr^2 P(x) + dx^2 Q(r)] R^2(r, x)}{P(x) Q(r) \Omega^2(r, x)} \ ,
\eeq
\beq
            A_\m dx^\m = - \frac{ e r + \om p x}{R^2(r,x)} \ dt \ + \ \left[ \frac{e \om r x^2 - p x r^2}{R^2(r,x)} + \hat{\om}_0 A_t(r,x)\right]  d\varphi \nn\ ,
\eeq
where
\bea
        Q(r) &=&  \hat{k} \om^2 + p^2 + e^2 -2mr + r^2 \e - \frac{2 n r^3 A}{\om} - \frac{r^4 \L}{3} - \hat{k} r^4 A^2   \ , \nn \\
        P(x) &=&  \hat{k} + \frac{2nx}{\om} - x^2 \e + 2 m x^3 A +x^4 \left[ - \frac{\L \om^2}{3} - A^2 \Big( p^2 + e^2 + \hat{k} \om^2 \Big) \right]  \nn  \ , \\
        R^2(r,x) &=& r^2 + \om^2 x^2 \ , \nn \\
        \Om(r,x) &=& 1 - A x r \ . \label{PD-ds2-fine}
\eea  
$m, \, \om, \, e, \, p, \, n, \, A, \, \hat{k}, \, \e, \, \hat{\om}_0, \, $ are integrating constants related to the mass, the electric and magnetic monopolar charges, the nut parameter, the acceleration, the topology and the asymptotic angular speed of a rotating observer. In the non-accelerating case $A=0$ we recover the topological Kerr-Newmann solutions (\ref{KN-topo})-(\ref{KN-Am}) for
\bea
          n &\to& k \ell - \frac{4\ell^3 \L}{3} + a^2 \left( \frac{k_7}{2\ell} + \frac{k^2 \ell \L}{3}  \right) \ , \nn \\
          \e &\to& k - \frac{\L}{3} a^2 k^2 -2\ell^2\L \ , \nn \\
          \hat{k} &\to& \frac{-k\ell^2+\ell^4\L+a^2 \big[1+k-k^2(1+\ell^2 \L) \big]}{\om^2}  \ , \nn \\
          \hat{\om}_0 &\to & \frac{\ell^2}{a} + \omega_0 \ , \hspace{1.9cm} \omega \to a \ , \nn \\
               x &\to & \frac{a x+\ell}{\omega} \ \nn . 
\eea
On the other hand, in the presence of the acceleration, the above map should be extended\footnote{For instance $ \e \to 1 - (a^2 - e^2 - p^2) A^2 -2\ell^2\L, \ \ k_7 = - \ell [k \ell + 2 a^2 m A +a^2 k \ell \l -\ell^3 \L \ell]/a^2  $ gives the accelerating black hole in the parametrization \cite{marcoa-equivalence}.}. \\
In any case, the generalised background, also in the presence of the acceleration, that is the natural accelerating extension of (\ref{KN-dwr})-(\ref{A-dwr}), is contained into the conjugation or the double Wick rotation (\ref{dwr}) of the solution (\ref{PD-ds2})-(\ref{PD-ds2-fine}).\\

\section{Conclusions}

In this article we have taken into consideration all the known exact, analytical and physical well behaved black hole solutions in four-dimensional general relativity (possibly coupled with standard electromagnetism). Metrics that are not necessarily asymptotically flat, but whose geometry does not present conical, or curvature singularities outside the event horizon. We have noticed that these spacetimes belong to a common family and that they can be comprehensively classified according to the properties of their background. This general class has an unexplored sector. We have identified the background characterising this unknown sector and we have built new black hole solutions whose background possesses these novel physical features. In this way we can provide a more detailed and rich black holes picture in the Einstein-Maxwell theory of gravitation. The last unexplored parameter stems from the conjugated general black hole metric of type D and it is directly related to the angular momentum in the black hole metric (before the conjugation). This integration constant has been so far overlooked probably because it does not directly descend from Lie-point transformations such as the electromagnetic charges or the NUT parameter, which are strictly connected to the Harrison and Ehlers symmetries of the Ernst's equations. Nevertheless the inverse scattering technique has been able to take into account also this extra parameter. This new rotational feature naturally can be superposed to the already known ones such as the Bonnor-Melvin and the swirling parameters. \\

These results point to a remarkable insight: basically {\it all the known legitimate axisymmetric and stationary black hole solutions in the Einstein-Maxwell theory are (accelerating) Kerr-Newman-NUT black hole embedded into the double Wick rotated (accelerating) Kerr-Newman-NUT background.} \\ 

All these metrics can be built thanks to the axisymmetric and stationary solution generating techniques. The inverse scattering method also suggests a physical interpretation for these backgrounds, as the gravitational and electromagnetic fields of far away black holes. This is coherent with the interpretation of the Bonnor-Melvin universe as a couple of Reissner-Nordstrom black holes at infinity \cite{Emparan:2001gm} or the expanding bubble of nothing as a couple of big Schwarzschild black holes that extend to spatial infinity \cite{bubble}.\\ 
Now the bigger open question is about the uniqueness or eventual further generalisation of these regular backgrounds. In general, according to the inverse scattering method, it would be possible to add extra solitons at infinity to add extra integrating constants. However it is not clear if these possible extra degrees of freedom might be reabsorbed in the previous ones.\\
What we have done here can be promptly adapted to other gravitational theories, closely related to general relativity, such as  minimally and conformally coupled scalar fields, non-linear sigma models or some Brans-Dicke and $f(R)$ theories, as noticed in \cite{marcoa-embedding} or \cite{marcoa-stationary}. \\

\paragraph{Acknowledgements}
{\small I would like to thank Riccardo Martelli and Omar Samuel Contin for numerous stimulating discussions on this subject. A Mathematica notebook containing several solutions presented in this article can be found in the arXiv source folder.}\\

\vspace{0.1cm}

\appendix

\section{About black holes in the Levi-Civita background}
\label{app:levi-bh}

Recently there have been some interest about black holes in a Levi-Civita background, Schwarzschild or Kerr with an asymptotic that is not Minkowskian but a special subclass of the Levi-Civita metric, see for instance \cite{maccallum}, \cite{kerr-LC}. As noticed in \cite{bh+BRBM}, these solutions belong to the more general class of black holes in the swirling background. The rationale is quite straightforward: Black hole in the Levi-Civita background can be generated from the (magnetic) inversion transformation of the Ernst equations, but the inversion transformation is contained into the (magnetic) Ehlers transformation, hence the black holes into the Levi-Civita background are contained into the Ehlers transformed Schwarzschild or Kerr black hole. Unfortunately misleading and incorrect literature is continuing to emerge ignoring this inclusion, therefore, in this appendix, we further clarify this point, explicitly verifying the relation between the swirling and Levi-Civita black holes, for some particular cases.\\
\paragraph{Kerr--Levi-Civita from Kerr-swirling} We want to prove that the inversion transformation of the Kerr black hole is contained into the the Ehlers transformed Kerr metric for a specific value of the parameter, $\jmath$, introduced by the Ehlers transformation. The Kerr metric deformed by a magnetic Ehlers transformation has been found in \cite{swirling} and describes a Kerr black hole embedded into a rotating background (dubbed swirling universe for its peculiar properties). Following the notation of \cite{swirling} (with $x=\cos \theta$), the metric can be cast in the following Lewis-Papapetrou metric
\beq
          ds^2 = f(d\varphi - \omega dt)^2 + \frac{1}{f} \left[ -\r^2dt^2 + \S (1-x^2) \left( \frac{dr^2}{\D_r} + \frac{dx^2}{1-x^2} \right) \right] \ ,
\eeq
with
\bea
            f(r,x) &=& \frac{\S (1-x^2) R^2}{(R^2+2\jmath m a \Xi x)^2 + \jmath^2 \S^2 (1-x^2)^2}  \ , \nn \\  
           \om(r,x) &=& \bar{\om}_0 - \frac{2mar}{\S} + \jmath \frac{4 x (r^3-a^2(m+(m-r)x^2))\D_r}{\S} \nn \\
                     &+& \jmath^2 2 a m \frac{(a^2-3r^2)(r^3+a^2(r+2m)) + x^2(r^3(x^2-6)+a^2(2m(3+x^2))-3r(2+x^2))\D_r}{\S} \ , \nn
\eea
where
\bea
            \D_r &=&  r^2 -2mr + a^2 \ ,  \hspace{3.4cm}  \Xi = r^2(x^2-3)-a^2(1+x^2) \ , \nn \\
            \r^2 &=&  \D_r (1-x^2)  \ ,    \hspace{3.9cm}  R  = r^2 + a^2 x^2   \ ,\nn \\
            \S   &=&  (r^2+a^2)^2 + a^2 \D_r (1-x^2) \ ,  \hspace{1.5cm}  \bar{\om}_0  =  \om_0 + - \frac{4c^2m^2(3a^4+16a^2m^2-16m^4)}{a^3}   \ .  \nn
\eea
To pick the leading order in the limiting procedure we rescale some  parameters and coordinates and choose the rotating constant gauge $\om_0$ as follows 
\beq
       m \to \frac{m}{\jmath} \ , \qquad a \to -\frac{a}{\jmath}  \ , \qquad t \to \frac{t}{\jmath} \ , \qquad \varphi \to \varphi \jmath^2  \ , \qquad r \to \frac{r}{\jmath} \ , \qquad \jmath \to \jmath^3 \ , \qquad \om_0 \to \frac{64m^4(m^2-a^2)}{a^3} \jmath^3 \ . \nn
\eeq
Then taking the limit for large swirling parameter $\jmath$ we get
\beq \label{GV-ds}
     ds^2 = \tilde{f} (d\varphi - \tilde{\omega} dt)^2 - \frac{\D_r(1-x^2)}{\tilde{f}} dt^2 + \frac{\e^{2\tilde{\g}}}{\tilde{f}} \left(\frac{dr^2}{\D_r} +\frac{dx^2}{1-x^2}\right) \ ,
\eeq
where
\bea \label{GV-f}
       \tilde{f} &=& \frac{R^2 (1-x^2) \S}{4m^2a^2x^2[a^2(1-x^2)^2+R^2(3-x^2)]^2 (1-x^2)^2 \S} \\
       \tilde{\om}   &=& - 2 m a \frac{(2a^2m-3a^2r+r^3)\D_r x^4 -6r(a^2+r^2)\D_r x^2 + (2a^2m+a^2r+r^3)(a^2-6mr-3r^2)}{\D_r a^2 x^2 +r(2a^2m+a^2r+r^3)} \ , \nn
\eea
which is precisely the solution obtained with the inversion transformation of the Kerr seed \cite{kerr-LC}, basically even in the same set of coordinates. \\
Actually using the gauge freedom typical of axisymmetric and stationary vacuum spacetimes in general relativity, $\S \to \S/\jmath^2$ we can alternatively get the same result easier, only by setting 
\beq
        a \to - a \ , \qquad  t \to \frac{t}{\jmath}\ , \qquad \varphi \to \varphi \jmath \ , \qquad  \om_0 \to \frac{64m^4(m^2-a^2)}{a^3} \jmath^2 \ ,
\eeq
before performing the limit.\\
So we have shown that the Kerr-Levi-Civita solution \cite{kerr-LC} is a subcase of the Kerr-swirling \cite{swirling}, for a specific value of the swirling parameter $\jmath$. %, since the first is restrained to a specific value of $\jmath$, their physical interpretation, in general, can differ. To explain it with a simple example set,  for instance, $ m = - 13 $ in the Schwarzschild metric you can not interpret it as a black hole any more neither it is not connected to Minkowski background (for finite spatial distances), as the full Schwarzschild metric. However all the physical and mathematical information of $ m = - 13 $ Schwarzschild metric is already contained in the general Schwarzschild solution. Similarly 
All the information of the Kerr-Levi-Civita metric is yet encoded, by construction, into the swirling Kerr solution.

\paragraph{Schwarzschild--Levi-Civita from Schwarzschild-swirling} Similarly the metric found in \cite{maccallum}, stems from the Schwarzschild in the swirling background \cite{swirling}. Clearly would be sufficient to switch off the angular momentum parameter $a$ in the above section, however for maximum clarity let us derive it directly from the swirling Schwarzschild metric. To get the Schwarzschild-Levi-Civita metric from the swirling Schwarzschild we have just to take the limit for $c=\infty$ of the swirling metric. To take the limit properly, the following rescaling of the some coordinate or parameters has to be considered
\beq
         m \to \frac{m}{\jmath} \ , \qquad \qquad \ t \to \frac{t}{\jmath} \ , \qquad \qquad r \to \frac{r}{\jmath} \ , \qquad \qquad \varphi \to \varphi \jmath^2 \ , \qquad \qquad \jmath \to \jmath^3 \ .
\eeq 
As in the rotating case a quicker path for the same result comes from taking advantage of the gauge freedom of the $\g$ functions, as in (\ref{curlin-melvin-schwarzschild-inizio}). In this case one needs to rescale only $t \to t/\jmath$ and $\varphi \to \varphi \jmath$.\\

\paragraph{Schwarzschild-swirling in an external gravitational background}

Another example where the large swirling parameter limit gives a solution obtained through the inversion transformation comes  from the Schwarzschild-swirling black hole embedded in an external gravitational field, recently found in eqs. (4.12), (4.14)-(4.15) of \cite{bh+BRBM}. This metric specialises into the Schwarzschild-Levi-Civita into the external gravitational background for $c=\infty$. To evaluate the metric in $c=\infty$ we preliminary set the gauge freedom by
\beq
          t \to \frac{t}{c} \ , \hspace{2cm} \varphi \to c \, \varphi \ , \hspace{2cm} \g_0 \to - \log c \ . \nn
\eeq   
The resulting spacetime coincides with the non accelerating case studied into section III.B of \cite{inv-GV}.\\
Moreover note that the Ehlers transformation (for big values of its Lie parameter) can remove the Bertotti-Robinson electromagnetic field surrounding the black hole for $c=\infty$. For instance if we consider the large swirling regime of the Schwarzschild-swirling-Bertotti-Robinson of appendix B of \cite{bh+BRBM} we get again the Schwarzschild-Levi-Civita into the external gravitational background, so the electromagnetic field vanished. However we note that at this scope the Harrison transformation has to be preferred, because it is more precise and gives more general and physical results. In fact the metric obtained from the Harrison transformation contains not only the Levi-Civita background but also the Minkowski ground state, moreover it is continuously connected to the seed.\\

%Finally note that, even though verifications explicitly performed here involved only black hole spacetimes, this is a general property independent of the physical interpretation. Solutions generated through the inversion transformation of a given seed stems from the Ehlers transformation of the same seed for large values of the Ehlers transformation; see next section for further details.

\subsection{Inversion, Ehlers and Harrison transformations}

It may be worth emphasizing that the above direct verifications are completely unnecessary because conceptually they stem just from the properties of the Lie-point transformations. In fact, as mentioned, the inversion transformation is not an independent symmetry of the Ernst equations, but it is contained in other transformations, such as the Ehlers transformation, see \cite{stephani-big-book} or footnote 5 of \cite{enhanced}. To put things into context we can say that the axisymmetric and stationary space of solutions of Einstein-Maxwell theory enjoys an eight parameter continuous group of transformations. These transformations allow us to move in the space of axisymmetric and stationary solutions of the theory, thus generating new solutions starting from old ones (called seed). These SU(2,1) symmetries of the system can be arranged into the following five Lie point transformations
\bea \label{su21-transf}
      (I)    && \Er \longrightarrow \Er' = \l \l^* \Er  \qquad \ \quad \qquad \ ,  \qquad \mathbf{\Phi} \longrightarrow  \mathbf{\Phi}' = \l \mathbf{\Phi} \quad , \nn \\
      (II)   && \Er \longrightarrow \Er' = \Er + i \ b \qquad  \ \ \ \qquad, \qquad \mathbf{\Phi} \longrightarrow  \mathbf{\Phi}' = \mathbf{\Phi} \quad ,  \nn \\
      (III)  && \Er \longrightarrow \Er' = \frac{\Er}{1+i\jmath\Er} \qquad \ \ \qquad , \qquad  \mathbf{\Phi} \longrightarrow  \mathbf{\Phi}' = \frac{\mathbf{\Phi}}{1+i \jmath\Er} \quad ,  \\
      (IV)   && \Er \longrightarrow \Er' = \Er - 2\b^*\mathbf{\Phi} - \b\b^* \ \ \ , \qquad \mathbf{\Phi} \longrightarrow  \mathbf{\Phi}' = \mathbf{\Phi} + \b  \quad ,  \nn \\
      (V)    && \Er \longrightarrow \Er' = \frac{\Er}{1-2\a^*\mathbf{\Phi}-\a\a^*\Er} \  , \quad  \ \ \mathbf{\Phi} \longrightarrow  \mathbf{\Phi}' = \frac{\mathbf{\Phi}+\a\Er}{1-2\a^*\mathbf{\Phi}-\a\a^*\Er}\ \quad  \nn
\eea
where the parameters $b, \jmath \in \mathbb{R}$ and $\a, \l,\b \in \mathbb{C}$. The most interesting transformations are the $(I)$ and the $(V)$
called the Ehlers transformation and the Harrison transformation, the $(I)$ can act as a duality transformation between the electric and the magnetic field, while the others can be reabsorbed by gauge and coordinate transformations, for more informations see \cite{stephani-big-book}, \cite{enhanced} or \cite{riccardo-tesi-bachelor}. \\
Composing the elements of this group of transformations it is possible to build others transformations. Of course we cannot expect that the composite transformation can extend the group and generate novel solutions with respect to the given set (\ref{su21-transf}). A well known discrete transformation that can be derived from the set (\ref{su21-transf}) is the inversion transformation \footnote{In vacuum subcase the inversion transformation is sometimes called Buchdahl transformation.}
\beq \label{inv}
         (inv)\hspace{2cm}    \bar{\Er} = \frac{1}{\Er} \ , \hspace{2cm} \bar{\Phi} =  \frac{\Phi}{\Er} \ . \nn
\eeq
It can be obtained in several ways, for instance by combining $(I) \circ(III) \circ(II) $ for $\jmath=1/b$ and $\l=i/b$ and then taking the limit for $ b \rightarrow \infty $. This point is even clearer if we consider the enhanced Ehlers transformation, as defined in \cite{enhanced} 
\beq \label{enhanced-ehlers}
            \Er \longmapsto \bar{\Er} = \frac{\Er + i \jmath}{1+i \jmath \Er} \ , \hspace{2cm} \Phi \longmapsto \bar{\Phi} =  \frac{\Phi(1 + i \jmath)}{1 + i \jmath \Er} \ , 
\eeq
which is a variation of the Ehlers transformation build\footnote{To obtain the enhanced Ehlers transformation just compose $(II) \circ (I) \circ (III)$, with $b=j$ and $\l = 1+ij$.} to preserve some asymptotic properties of the asymptotically flat seeds. The enhanced Ehlers transformation is built from the standard Ehlers transformation combined with some gauge transformations, therefore its effect on the seed metric is physically equivalent. As noted in \cite{PD-NUTs}, the large $\jmath$ limit of the enhanced Ehlers transformation (\ref{enhanced-ehlers}) is the inversion transformation (\ref{inv}). That is the reason why we are taking the large $\jmath$ limit to get the Kerr-Levi-Civita metric from the Kerr swirling solution\footnote{Note that if, contrary to the standard convections, we define $\jmath=1/c$, the enhanced Ehlers transformation becomes the identity for $c=\infty$ and it would suggest us to take the $c=0$ to get Kerr-Levi-Civita from Kerr swirling. So, in general, depending on how we parametrise the Ehlers transformation, we would have a different limiting value for the swirling parameter to obtain the Levi-Civita background.}. \\
So, by construction, the Ernst fields of the transformed seed according to the inversion transformation are a special case of the ones obtained with the Ehlers transformation (both the standard and the enhanced), up to gauge transformations. Since the Ernst complex potentials determine univocally the metric and electromagnetic potential solution, up to gauge transformations, also at the metric level, solutions obtained from the inversion transformation are a subclass of the ones obtained from the Ehlers transformation acting on the same seed.  \\
This is true for any seed not just for the explicit examples we explicitly treated above, also non-black hole spacetimes. So the inversion transformation can not really provide new solutions with respect to the Ehlers transformation. Moreover, generating solution with a Levi-Civita background from already known swirling ones is faster than using the symmetries of the Ernst equations. But it is not just a matter of generality. The Ehlers, as other non-trivial Lie-point transformations, contrary to the inversion transformation, which is discrete, brings in the initial solution a new integrating constant continuously deforming the seed. This means that the Lie-point transformations can generate solutions as close as necessary to the seed to fit the actual phenomenology, still carrying new physical features. \\

\paragraph{Schwarzschild-Levi-Civita from Schwarzschild-Bonnor-Melvin} The fact that in section \ref{schwarzschild-kerrling} we have obtained the Schwarzschild-Levi-Civita black hole as a limit of the curling Schwarzschild metric makes we question if the superposition of black holes (or any metric) into the Levi-Civita background can stem from other spacetimes apart from the black holes solutions in the curling or in the swirling background. Maybe it is a common feature with a larger class that share similar Levi-Civita asymptotic. We know that also black holes embedded into the Bonnor-Melvin universe have a similar fall off for large radial distances, so it could be interesting to verify if this family of solutions can also contain Schwarzschild-Levi-Civita or Kerr-Levi-Civita.\\
At this scope consider the Schwarzschild solution embedded into the external magnetic field of Bonnor-Melvin, as written is eq. (\ref{melvin-bh-ds})-(\ref{melvin-bh-A}) and rescale some coordinates and parameters to execute the limit properly  
\beq
         t \to t \, \frac{4}{b^2} \ , \qquad \varphi \to \varphi \, \frac{b^2}{4} \ , \qquad \g_0 \to \log \frac{4}{b^2} \ , \qquad A_{\varphi_0} \to \frac{2}{b} \ . \nn
\eeq
Then taking the limit for $b\to\infty$ you get again exactly the Schwarzschild-Levi-Civita solution (\ref{sch-LC-ds}).

\paragraph{Kerr-Newman-Levi-Civita from Kerr-Newman-Bonnor-Melvin} Similarly to the above cases we can obtain the rotating and charged black hole with Levi-Civita background as a special case of the Kerr-Newman-Bonnor-Melvin solution, build in \cite{ernst-wild}. Thus, consider it in the notation\footnote{Note that the parameter denoting the intensity of the external Bonnor-Melvin electromagnetic field, $B$ of \cite{magnetised-kerr-cft}, is relabelled here as $b$ to homogenise the notation, and avoid confusion with the Bertotti-Robinson field denoted by $B$ in this article.} of \cite{magnetised-kerr-cft} and take the leading order of large $b$ regime for the solution, this automatically select the needed rescaling of the Killing coordinates and the gauge constants $\g_0$, $A_{t_0}$ and $A_{\varphi_0}$ as
\beq
        t \to t \, \frac{4}{b^2} , \qquad \quad \varphi \to \varphi \, \frac{4}{b^2}  \ , \qquad \quad \g_0 \to \log \frac{4}{b^2} \ , \qquad \quad A_{t_0} \to \frac{3}{2} am^2b^3 \ , \nn
\eeq
\beq
           A_{t_0} \to \frac{3}{2} am^2b^3 \ ,  \qquad \qquad A_{\varphi_0} \to \left[e^4\left(\frac{1}{8} -\frac{1}{128a^2m^2} -\frac{1}{8} +2a^2m^2 \right) \right] b^3 \ ,\nn
\eeq
while the other arbitrary constants, such as the rotational speed at infinity or the dilatation of the azimuthal angle are chosen just to make contact with the above uncharged metric
\beq
        \D_\varphi = \frac{1}{16a^2m^2} \ , \qquad \qquad  \om_0 = \frac{3am^2b^4}{4} \ ,   \nn
\eeq
even though, to avoid conical singularities, $\D_\varphi$ should take a unitary value. \\
The resulting metric is precisely the generalisation of the Kerr-Levi-Civita (\ref{GV-ds})-(\ref{GV-f}) in the presence of monopolar electric charge, so the Kerr-Newman-Levi-Civita. We leave this solution into a Mathematica notebook\footnote{The Mathematica file can be found between the arxiv source files.}, for the interested reader. From this procedure it can be appreciated as spacetimes in the Levi-Civita background can be obtained in a simple way, without using the symmetry transformations of the Ernst equations or passing through the complex Ernst potential, in the case that these solution are already available, since they have generated through the Ehlers, the Harrison or other generating techniques (as the solution of section \ref{sec:bh+bacground}).
\\

\paragraph{The Schwarzschild-Levi-Civita into the external gravitational background from Schwarzschild Bertotti-Robinson-Bonnor-Melvin} The Schwarzschild-Levi-Civita metric embedded in the external gravitational background described in \cite{static-typeI-bh}, is also contained into the Schwarzschild Bertotti-Robinson-Bonnor-Melvin, i.e. the solution in (2.18)-(2.20) of \cite{bh+BRBM}, for $b=\infty$. To verify explicitly, before evaluate the metric in $b=\infty$, set the gauge freedom as follows 
\beq
             t \to t \, \frac{4}{b^2} \ , \qquad \quad \varphi \to \varphi \, \frac{4}{b^2}  \ , \qquad \quad \g_0 \to \log \frac{4}{b^2} \ .
\eeq
Note that automatically the electromagnetic field vanishes. Exactly the same procedure works well in the presence of acceleration. Therefore the accelerating Schwarzschild-Levi-Civita into the external gravitational background is a special case of the accelerating Schwarzschild into the Bertotti-Robinson-Bonnor-Melvin electromagnetic field, there is no need to generate it twice, as they were independent, as done in \cite{inv-GV}. \\

Now that the nature of these solutions obtained with the inversion transformation has been elucidated, it is clear how to write fundamentally any generalisation. For instance, since the accelerating Kerr-Newman-Levi-Civita is a subcase of the swirling accelerating Kerr-Newman solution of \cite{removal}, if we want to write the metric, we have just to
rescale $t\to t/\jmath$, $\varphi \to \varphi/\jmath$ and set $ \g_0 \to - \log \jmath\,$ before evaluating the swirling solution for large values of the swirling parameter $\jmath$, the full solution is provided in the Mathematica notebook for the interested reader.\\

\paragraph{Inversion from the Harrison transformation} It is interesting to understand why it is possible to recover solutions in the Levi-Civita background also as a special case of metrics in the Bonnon-Melvin background. From a solution generating perspective the motivation relays in the fact that the inversion transformation is a particular subcase, not only of the Ehlers, but also of the Harrison transformation.\\
Indeed by composing the real Harrison transformation with two gauge transformation, and taking the limit for $\b \to \infty$ we get the inversion transformation:
\beq \label{harry-3}
             (I) \circ (V) \circ (IV) \bigg|_{\substack{ \l = 1/\b \\ \a = 1/\b}} \qquad  \xlongrightarrow[\b \to \infty] \quad \hspace{0.3cm} (inv) 
\eeq

In case we want to get the inversion transformation without applying the limiting procedure, it is possible to to compose the above combination (\ref{harry-3}) with another real $(IV)$ transformation
\beq
                    (I) \circ (IV) \circ (V) \circ (IV) \bigg|_{\substack{ \l = 1/\b \\ \a = 1/\b}}  = (inv) \ .
\eeq
However, especially at the metric level, the limit representation seems more insightful and practical to reproduce the inversion transformation result, from a given solution which already possesses a swirling or Bonnor-Melvin background. \\
Of course this shortcut to reproduce the action of the inversion transformation from the Ehlers or the Harrison transformation is not limited to the magnetic version of the transformation seen above, but it is possible to use it to reproduce the action of the electric inversion transformation from metrics that possess electromagnetic monopolar charges or the NUT parameter: Reissner-Nordstrom or Taub-NUT.
For instance the Schwarzschild-NUT metric in the large NUT parameter regime (large $\ell$) gives\footnote{To verify it explicitly just choose $\g_0=1/\ell$, rescale the coordinates and the parameters as follows: $\varphi \to \varphi/\ell, \ r \to \ell r, \ m \to \ell m, \ \ell \to \sqrt{\ell}$ before taking the $\ell \to \infty$ limit. Incidentally in this case the same limit to the Schwarzschild spacetime can be obtained also for $\ell \to 0$ from the Schwarzschild-NUT, but this is a peculiarity of the seed considered in this simple example, related to the  uniqueness of asymptotically flat black holes.} the electrically\footnote{Regarding the electric and magnetic symmetries transformations see \cite{swirling}, \cite{riccardo-tesi-bachelor}.} inverted Schwarzschild metric (which is again the Schwarzschild black hole).
%As done for the (III) symmetry, can be interesting to take into consideration the enhanced version if the Harrison transformation to make even more apparent the relation with the inversion transformation 
\\


\vspace{1cm}


\begin{thebibliography}{99}

\bibitem{kerr}
R.~P.~Kerr,
{\it ``Gravitational field of a spinning mass as an example of algebraically special metrics''},
\href{https://doi.org/10.1103/PhysRevLett.11.237}{Phys. Rev. Lett. \textbf{11} (1963), 237-238}.
%doi:10.1103/PhysRevLett.11.237
%2847 citations counted in INSPIRE as of 11 Dec 2025

\bibitem{newman}
E.~T.~Newman, E.~Couch, K.~Chinnapared, A.~Exton, A.~Prakash and R.~Torrence,
{\it ``Metric of a Rotating, Charged Mass''},
\href{https://doi.org/10.1063/1.1704351}{J. Math. Phys. \textbf{6} (1965), 918-919}.
%doi:10.1063/1.1704351
%1046 citations counted in INSPIRE as of 11 Dec 2025

\bibitem{chandrasekhar}
S.~Chandrasekhar,
{\it ``The mathematical theory of black holes''}, Clarendon Press, Oxford (1985)
%373 citations counted in INSPIRE as of 28 Aug 2025

\bibitem{swirling}
 M.~Astorino, R.~Martelli and A.~Vigan\`o,
  {\it ``Black holes in a swirling universe''},
   \href{https://doi.org/10.1103/PhysRevD.106.064014}{Phys. Rev. D \textbf{106} (2022) no.6, 064014};
    \href{https://arxiv.org/pdf/2205.13548}{\tt [arXiv:2205.13548 [gr-qc]]}

\bibitem{removal}
 M.~Astorino,
  {\it ``Removal of conical singularities from rotating C-metrics and dual CFT entropy''}
   \href{https://doi.org/10.1007/JHEP10(2022)074}{JHEP \textbf{10} (2022), 074};
    \href{https://arxiv.org/pdf/2207.14305}{\tt [arXiv:2207.14305 [gr-qc]]}

\bibitem{bonnor}
W.~B.~Bonnor,
{\it ``Static Magnetic Fields in General Relativity''},
\href{https://doi.org/10.1088/0370-1298/67/3/305}{Proc. Roy. Soc. Lond. A \textbf{67} (1954) no.3, 225}
%73 citations counted in INSPIRE as of 06 Apr 2026
    

\bibitem{melvin}
M.~A.~Melvin,
{\it ``Pure magnetic and electric geons''},
\href{https://doi.org/10.1016/0031-9163(64)90801-7}{Phys. Lett. \textbf{8} (1964), 65-70}
%442 citations counted in INSPIRE as of 06 Apr 2026

\bibitem{ernst-wild}  
  F.~J.~Ernst and W. Wild,
  {\it ``Kerr black holes in a magnetic universe''},
   \href{https://doi.org/10.1063/1.522875}{J.\ Math.\ Phys.{\bf 17} (1976) 182}

\bibitem{kerr-bertotti}
J.~Podolsky and H.~Ovcharenko,
{\it ``Kerr black hole in a uniform magnetic field: An exact solution''},
\href{https://doi.org/10.1103/rfgv-ybz5}{Phys. Rev. Lett. \textbf{135} (2025) no.18, 181401};\
%doi:10.1103/rfgv-ybz5
\href{https://arxiv.org/pdf/2507.05199}{\tt [arXiv:2507.05199 [gr-qc]]}
%5 citations counted in INSPIRE as of 17 Aug 2025

\bibitem{Ovcharenko:2025cpm}
H.~Ovcharenko and J.~Podolsk{\'y},
{\it ``New class of rotating charged black holes with nonaligned electromagnetic field''},
\href{https://doi.org/10.1103/8wkz-th6v}{Phys. Rev. D \textbf{112} (2025) no.6, 064076};
\href{https://arxiv.org/pdf/2508.04850}{\tt [arXiv:2508.04850 [gr-qc]]}
%19 citations counted in INSPIRE as of 02 Apr 2026

\bibitem{uniqueness}
M.~Heusler,
{\it ``Black Hole Uniqueness Theorems''}, 1996,
\href{https://doi.org/10.1017/cbo9780511661396}{ISBN 978-0-511-66139-6}.
%50 citations counted in INSPIRE as of 11 Dec 2025

\bibitem{malek}
A.~Coll{\'e}aux, I.~Kol{\'a}{\v{r}} and T.~M{\'a}lek,
{\it ``Double Wick rotations between symmetries of Taub-NUT, near-horizon extreme Kerr, and swirling spacetimes''},
\href{https://arxiv.org/pdf/2509.22309}{\tt [arXiv:2509.22309 [gr-qc]]}.
%0 citations counted in INSPIRE as of 12 Dec 2025

\bibitem{tesi-riccardo}
R.~Martelli,
{\it ``Charged Black hole solutions in general relativity from inverse scattering''},\
\href{https://github.com/Riccardo-Martelli/Master-thesis/blob/main/Tesi_Magistrale_Riccardo_Martelli_02371A_final.pdf}{Master thesis, Universit\`a degli studi di Milano (2024).}
%0 citations counted in INSPIRE as of 11 Dec 2025

\bibitem{tesi-omar}
O. S. Contin
{\it ``Exact and analytical black holes embedded into back-reacting rotating spacetimes in general relativity''},\
Master thesis, Universit\`a degli studi di Milano (2024).

\bibitem{charging}
M.~Astorino,
{\it ``Charging axisymmetric space-times with cosmological constant''},
\href{https://doi.org/10.1007/JHEP06(2012)086}{JHEP \textbf{06} (2012), 086};
\href{https://arxiv.org/pdf/1205.6998.pdf}{\tt [arXiv:1205.6998 [gr-qc]]}
%35 citations counted in INSPIRE as of 27 Sep 2022

\bibitem{ortin}
N.~Alonso-Alberca, P.~Meessen and T.~Ortin,
{\it ``Supersymmetry of topological Kerr-Newman-Taub-NUT-AdS space-times''},
\href{https://doi.org/10.1088/0264-9381/17/14/312}{Class. Quant. Grav. \textbf{17} (2000), 2783-2798}; 
%doi:10.1088/0264-9381/17/14/312
\href{https://arxiv.org/pdf/hep-th/0003071}{\tt [arXiv:hep-th/0003071 [hep-th]]}
%102 citations counted in INSPIRE as of 10 Nov 2025

\bibitem{klemm-erratum}
D.~Klemm, V.~Moretti and L.~Vanzo,
{\it ``Rotating topological black holes''},
\href{https://doi.org/10.1103/PhysRevD.60.109902}{Phys. Rev. D \textbf{57} (1998), 6127-6137 \ 
[erratum: Phys. Rev. D \textbf{60} (1999), 109902]}; 
%doi:10.1103/PhysRevD.60.109902
\href{https://arxiv.org/pdf/gr-qc/9710123}{\tt [arXiv:gr-qc/9710123 [gr-qc]]}
%130 citations counted in INSPIRE as of 10 Nov 2025

\bibitem{alekseev-inverse}
G.~A.~Alekseev,
{\it ``N-solitons solutions of the Einstein-Maxwell equations''},
Pisma Zh. Eksp. Teor. Fiz. \textbf{32} (1980), 301-303; 
\href{http://jetpletters.ru/ps/1426/article_21693.pdf}{JETP Lett. \bf{32}, 277 (1980)}
%17 citations counted in INSPIRE as of 24 Nov 2025

\bibitem{belinski-book}
 V. Belinski, E. Verdaguer, \href{https://doi.org/10.1017/CBO9780511535253}{\it ``Gravitational solitons''}, Cambridge, Cambridge Univ. Press, 2001.

\bibitem{belinski-19}
V.~A.~Belinski,
{\it ``On the black holes in external electromagnetic fields''},
\href{https://arxiv.org/pdf/1912.03964}{\tt [arXiv:1912.03964 [gr-qc]]}
%0 citations counted in INSPIRE as of 24 Nov 2025

\bibitem{marcoa-equivalence}
M.~Astorino,
{\it ``Equivalence principle and generalised accelerating black holes from binary systems''},
\href{https://doi.org/10.1103/PhysRevD.109.084038}{Phys. Rev. D \textbf{109} (2024) no.8, 8};
\href{https://arxiv.org/pdf/2312.00865.pdf}{\tt [arXiv:2312.00865 [gr-qc]]}.

\bibitem{harrison}
B. Kent Harrison, {\it ``New Solutions of the Einstein‐Maxwell Equations from Old''}, 
\href{https://doi.org/10.1063/1.1664508}{J. Math. Phys. 9 (11): 1744–1752, (1968)}

\bibitem{illy}
M.~Illy,
{\it ``Accelerated Reissner-Nordstrom black hole in a swirling, magnetic universe''},
Bachelor thesis, Universit\`a degli studi di Milano (2023);
\href{https://arxiv.org/pdf/2312.14995}{\tt [arXiv:2312.14995 [gr-qc]]}
%6 citations counted in INSPIRE as of 11 Dec 2025

\bibitem{adriano-andrea}
A.~Di Pinto, S.~Klemm and A.~Vigan{\`o},
{\it ``Kerr-Newman black hole in a Melvin-swirling universe''},
\href{https://doi.org/10.1007/JHEP06(2025)150}{JHEP \textbf{06} (2025), 150}, 
%doi:10.1007/JHEP06(2025)150
\href{https://arxiv.org/pdf/2503.07780}{\tt[arXiv:2503.07780 [gr-qc]]}
%4 citations counted in INSPIRE as of 17 Aug 2025

\bibitem{enhanced}
M.~Astorino,
{\it ``Enhanced Ehlers Transformation and the Majumdar-Papapetrou-NUT Spacetime''},
\href{https://doi.org/10.1007/JHEP01(2020)123}{JHEP \textbf{01} (2020), 123} ; 
%doi:10.1007/JHEP01(2020)123
\href{https://arxiv.org/pdf/1906.08228}{\tt [arXiv:1906.08228 [gr-qc]]}.
%3 citations counted in INSPIRE as of 17 Jan 2022

\bibitem{ernst-magnetic} 
  F.~J.~Ernst,
   {\it ``Black holes in a magnetic universe''},
    \href{https://doi.org/10.1063/1.522781}{J.\ Math.\ Phys.\  {\bf 17}, no. 1, 54 (1976).}

\bibitem{ernst-remove}
  F.~J.~Ernst,
   {\it ``Removal of the nodal singularity of the C-metric''}, 
    \href{http://scitation.aip.org/content/aip/journal/jmp/17/4/10.1063/1.522935}{J. Math. Phys. {\bf 17}, 515 (1976)}.

\bibitem{pair-rotating}
  M.~Astorino,
  {\it ``Pair Creation of Rotating Black Holes''},
  \href{https://doi.org/10.1103/PhysRevD.89.044022}{Phys. Rev. D \textbf{89} (2014) no.4, 044022};
  %doi:10.1103/PhysRevD.89.044022
  \href{https://arxiv.org/pdf/1312.1723.pdf}{\tt [arXiv:1312.1723~[gr-qc]]}.
  %9 citations counted in INSPIRE as of 20 May 2022

\bibitem{swirling-binary}
M.~Astorino and M.~Torresan,
{\it ``Rotating and swirling binary black hole system balanced by its gravitational spin-spin interaction''},
\href{https://doi.org/10.1103/3954-766n}{Phys. Rev. D \textbf{112} (2025) no.6, 6};\
%doi:10.1103/3954-766n
\href{https://arxiv.org/pdf/2502.08706}{\tt [arXiv:2502.08706 [gr-qc]]}.

\bibitem{levi-civita-BR}
T. Levi-Civita,
{\it ``Realt\`a fisica di alcuni spazi normali del Bianchi''}, Rend. Acc. Lincei, 26:519–531 (1917). \ {\it ``Republication of: The physical reality of some normal spaces of Bianchi''};
\href{https://doi.org/10.1007/s10714-011-1188-4}{Gen Relativ Gravit 43, 2307–2320 (2011)}.

\bibitem{garcia-alekseev}
G.~A.~Alekseev and A.~A.~Garcia,
{\it ``Schwarzschild black hole immersed in a homogeneous electromagnetic field''},
\href{https://doi.org/10.1103/PhysRevD.53.1853}{Phys. Rev. D \textbf{53} (1996), 1853-1867}.
%doi:10.1103/PhysRevD.53.1853
%19 citations counted in INSPIRE as of 01 Aug 2025

\bibitem{alekseev-RN-bertotti}
G.~A.~Alekseev,
{\it ``Charged black hole accelerated by spatially homogeneous electric field of Bertotti-Robinson (AdS2 x S2) space-time''},
\href{https://arxiv.org/pdf/2511.06082}{\tt [arXiv:2511.06082 [gr-qc]]}.
%1 citations counted in INSPIRE as of 11 Dec 2025

\bibitem{bh+BRBM}
M.~Astorino,
{\it ``Black holes in the external Bertotti-Robinson-Bonnor-Melvin electromagnetic field''},
\href{https://doi.org/10.1103/c5lw-53yd}{Phys. Rev. D \textbf{112} (2025) no.10, 104077};
%doi:10.1103/c5lw-53yd
\href{https://arxiv.org/pdf/2508.12908}{\tt [arXiv:2508.12908 [gr-qc]]}.
%4 citations counted in INSPIRE as of 10 Dec 2025

\bibitem{static-typeI-bh}
M.~Astorino,
{\it ``Static hairy black hole in 4D General Relativity''}, (2025); 
\href{https://doi.org/10.6084/m9.figshare.30081115.v1}{\tt https://doi.org/10.6084/m9.figshare.30081115.v1};
\href{https://doi.org/10.1103/yz86-wc3g}{Phys. Rev. D \textbf{113} (2026), 024047};
\href{https://arxiv.org/pdf/2601.16254}{\tt [arXiv:2601.16254 [gr-qc]]}

\bibitem{witten}
E.~Witten,
{\it ``Instability of the Kaluza-Klein Vacuum''},
\href{https://doi.org/10.1016/0550-3213(82)90007-4}{Nucl. Phys. B \textbf{195} (1982), 481-492}
%doi:10.1016/0550-3213(82)90007-4
%479 citations counted in INSPIRE as of 02 Oct 2021

\bibitem{horowitz-bubble-baths}
O.~Aharony, M.~Fabinger, G.~T.~Horowitz and E.~Silverstein,
{\it ``Clean time dependent string backgrounds from bubble baths''}, \href{https://doi.org/10.1088/1126-6708/2002/07/007}{
JHEP \textbf{07} (2002), 007};
%doi:10.1088/1126-6708/2002/07/007
\href{https://arxiv.org/abs/hep-th/0204158}{\tt [hep-th/0204158]}.
%132 citations counted in INSPIRE as of 02 Oct 2021

\bibitem{bubble}
 M.~Astorino, R.~Emparan and A.~Vigan\`o,
 {\it ``Bubbles of nothing in binary black holes and black rings, and viceversa''},
  \href{https://doi.org/10.1007/JHEP07(2022)007}{JHEP \textbf{07} (2022), 007};
   \href{https://arxiv.org/pdf/2204.09690.pdf}{\tt [arXiv:2204.09690 [hep-th]]}.

\bibitem{kerr+bubble}
M.~Astorino,
{\it ``Kerr Black Holes in an Expanding Bubble''},
\href{https://arxiv.org/pdf/2507.22114.pdf}{\tt [arXiv:2507.22114 [gr-qc]]}


\bibitem{emparan-reall}
 R.~Emparan and H.~S.~Reall,
  {``Generalized Weyl solutions''}, 
   \href{https://doi.org/10.1103/PhysRevD.65.084025}{Phys. Rev. D \textbf{65} (2002), 084025}; 
    \href{https://arxiv.org/pdf/hep-th/0110258.pdf}{\tt [arXiv:hep-th/0110258 [hep-th]]}.
     %280 citations counted in INSPIRE as of 07 Oct 2023

\bibitem{mostgeneral}
M.~Astorino,
{\it ``Most general type-D black hole and the accelerating Reissner-Nordstrom-NUT-(A)dS solution''},
\href{https://doi.org/10.1103/PhysRevD.110.104054}{Phys. Rev. D \textbf{110} (2024) no.10, 104054};
%doi:10.1103/PhysRevD.110.104054
\href{https://arxiv.org/pdf/2404.06551.pdf}{\tt [arXiv:2404.06551 [gr-qc]]}

\bibitem{new-look-PD}
J.~B.~Griffiths and J.~Podolsky,
{\it ``A New look at the Plebanski-Demianski family of solutions''},
\href{https://doi.org/10.1142/S0218271806007742}{Int. J. Mod. Phys. D \textbf{15} (2006), 335-370};
 \href{https://arxiv.org/pdf/gr-qc/0511091.pdf}{\tt [arXiv:gr-qc/0511091 [gr-qc]]}
%266 citations counted in INSPIRE as of 24 Mar 2026

\bibitem{PD-revisited}
H.~Ovcharenko, J.~Podolsky and M.~Astorino,
{\it ``Revisiting black holes of algebraic type D with a cosmological constant''},
\href{https://doi.org/10.1103/PhysRevD.111.084016}{Phys. Rev. D \textbf{111} (2025) no.8, 084016}; 
%doi:10.1103/PhysRevD.111.084016
\href{https://arxiv.org/pdf/2501.07537}{\tt [arXiv:2501.07537 [gr-qc]]}
%15 citations counted in INSPIRE as of 24 Mar 2026

\bibitem{Emparan:2001gm}
R.~Emparan and M.~Gutperle,
{\it ``From p-branes to fluxbranes and back''},
\href{https://doi.org/10.1088/1126-6708/2001/12/023}{JHEP \textbf{12} (2001), 023};
%doi:10.1088/1126-6708/2001/12/023
\href{https://arxiv.org/pdf/hep-th/0111177}{\tt [arXiv:hep-th/0111177 [hep-th]]}.
%53 citations counted in INSPIRE as of 29 Jan 2025


\bibitem{maccallum}
M.~M.~Akbar and M.~A.~H.~MacCallum,
{\it ``Static Axisymmetric Einstein Equations in Vacuum: Symmetry, New Solutions and Ricci Solitons''},
\href{https://doi.org/10.1103/PhysRevD.92.063017}{Phys. Rev. D \textbf{92} (2015) no.6, 063017};
%doi:10.1103/PhysRevD.92.063017
\href{https://arxiv.org/pdf/1508.05196}{\tt [arXiv:1508.05196 [gr-qc]]}

\bibitem{kerr-LC}
      J.~Barrientos, A.~Cisterna, M.~Hassaine, K.~M{\"u}ller and K.~Pallikaris,
        {\it ``A new exact rotating spacetime in vacuum: The Kerr{\textendash}Levi-Civita spacetime''},
         \href{https://doi.org/10.1016/j.physletb.2025.140035}{Phys. Lett. B \textbf{871} (2025), 140035};
          \href{https://arxiv.org/pdf/2506.07166}{\tt [arXiv:2506.07166 [gr-qc]]}


\bibitem{stephani-big-book}
     H.~Stephani, D.~Kramer, M.~A.~H.~MacCallum, C.~Hoenselaers and E.~Herlt,
       {``Exact solutions of Einstein's field equations''}, \ 
        \href{https://doi.org/10.1017/CBO9780511535185}{\tt [doi:10.1017/CBO9780511535185]}

\bibitem{PD-NUTs}
  M.~Astorino and G. Boldi,
     {\it ``Plebanski-Demianski goes NUTs (to remove the Misner string)''},
      \href{https://doi.org/10.1007/JHEP08(2023)085}{JHEP \textbf{08} (2023), 085};
       \href{https://arxiv.org/pdf/2305.03744.pdf}{\tt [arXiv:2305.03744 [gr-qc]]}

\bibitem{riccardo-tesi-bachelor}
         R.~Martelli,
         {\it ``The Action of the Axisymmetric and Stationary Symmetry Group of  General Relativity on a Static Black Hole''},
         \href{https://inspirehep.net/files/199a0153d1352cc4859a4f708d3799aa}{Bachelor thesis, Universit\`a degli studi di Milano (2022).},
%3 citations counted in INSPIRE as of 15 Mar 2026

\bibitem{inv-GV}
J.~Barrientos, A.~Cisterna, A.~D{\'\i}az and K.~M{\"u}ller,
{\it ``From Bertotti--Robinson to Vacuum: New Exact Solutions in General Relativity via Harrison and Inversion Symmetries''},
\href{https://arxiv.org/pdf/2602.17581v2}{\tt [arXiv:2602.17581 [gr-qc]]}

\bibitem{magnetised-kerr-cft}
M.~Astorino,
{ \it ``Magnetised Kerr/CFT correspondence''},
\href{https://doi.org/10.1016/j.physletb.2015.10.017}{Phys. Lett. B \textbf{751} (2015), 96-106};
\href{https://arxiv.org/pdf/1508.01583}{\tt [arXiv:1508.01583 [hep-th]]}.
%31 citations counted in INSPIRE as of 23 Jun 2022

\bibitem{marcoa-embedding}
M.~Astorino,
{\it ``Embedding hairy black holes in a magnetic universe''},
\href{https://doi.org/10.1103/PhysRevD.87.084029}{Phys. Rev. D \textbf{87} (2013) no.8, 084029} ;
\href{https://arxiv.org/pdf/1301.6794}{\tt [arXiv:1301.6794 [gr-qc]]}.
%21 citations counted in INSPIRE as of 21 May 2022


\bibitem{marcoa-stationary}
M.~Astorino,
{\it ``Stationary axisymmetric spacetimes with a conformally coupled scalar field''},
\href{http://doi.org/10.1103/PhysRevD.91.064066}{Phys. Rev. D \textbf{91} (2015), 064066} ;
\href{https://arxiv.org/pdf/1412.3539}{\tt [arXiv:1412.3539 [gr-qc]]}.



%%%%%%%%%%%%%%%%%%%%%%%%%%%%%%%%%%%%%%%%%%%%%%%%%%%%%%%%%%%%%%%%%%%%%%%%%%%%%%%%%%%%%%%%%%%%%%%%%%%%%%%%%%%%%%%%%%%%%%



\end{thebibliography}
\end{document}